\documentclass[12pt]{article}
 \usepackage{setspace}
\usepackage{amsmath}
\usepackage{amsfonts}
\usepackage{amssymb}
\usepackage{graphicx,color}
\usepackage{capt-of}
\usepackage[left=1in, top=1in, right=1in,bottom=1in, head=0.5in,foot=0.5in]{geometry}
\usepackage{amsthm}
\usepackage{extramarks}
\usepackage{titlesec}
\usepackage{rotating}
\usepackage{colortbl}
\usepackage{multirow}
\usepackage[sort&compress]{natbib}

\numberwithin{equation}{section}
\usepackage{tabularx}

\everymath{\displaystyle}
\usepackage{enumerate}
\usepackage{epstopdf}
\begin{document}

\author{\footnotesize Tracy Sweet$^1$ and Selena Wang$^2$\\
\footnotesize $^1$Department of Human Development and Quantitative Methodology, University of Maryland\\
\footnotesize  $^2$ Department of Biostatistics and Health Data Science, Indiana University School of Medicine}

\title{Network Science in Psychology}

\maketitle

\begin{abstract}
Social network analysis can answer research questions such as why or how individuals interact or form relationships and how those relationships impact other outcomes. Despite the breadth of methods available to address psychological research questions, social network analysis is not yet a standard practice in psychological research. To promote the use of social network analysis in psychological research, we present an overview of network methods, situating each method within the context of research studies and questions in psychology. 
\end{abstract}



\doublespacing
\section{Introduction}

For researchers interested in relationships or interactions, analyzing social network data is an obvious way to understand how individuals interact. Social network data -- also referred to as relational data in some fields -- refer to data about the presence or value of relationships among a set of individuals. Common examples of social networks include a friendship network among a group of children; a co-authorship network among a group of social psychologists; and collaboration network among researchers in a university.

Social network data take an unusual format; rather than just the rectangular array of $n \times p$ data, with $n$ rows of observations on $p$ covariates, a network consists of $n(n-1)$ network tie values, often represented as an $n \times n$ adjacency matrix, and then supplemented with an $n \times p$ array of data and sometimes with an additional $n(n-1)\times p $ array of data, which include individual-level and dyad-level attributes respectively. Network data can also be challenging to collect; data collection via survey generally involves asking respondents to identify and/or quantify their relationships and list individuals by name. 

Despite these additional levels of complexity, research questions that can be addressed with social network data are unique and cannot be achieved with other means. Further, network analytical methods can unearth novel research questions and enrich the scientific discussions surrounding many domains in psychology. 
In some domains, social network analysis has seen a surge in popularity, and studies about peer effects, social influence, and homophily are becoming standard. With additional methodological training and collaborations between substantive and methodological researchers, we argue that opportunities are abundant for new scientific frontiers.

The purpose of this paper is provide psychologists with a survey of social network analysis methods; unlike existing surveys on this topic, we emphasize the applicability of network methods in psychology, and our goal is to illustrate how these tools can enhance research. 
To that end, in most sections, we include an example from a psychological study and discuss how the study can be extended or augmented with social network analysis. Our goal is not to criticize existing work as we selected top scientific papers as exemplars of current research. Rather, our goal is to show how social network analysis can answer complementary research questions and provide additional knowledge.

In the current psychology literature, the term ``network'' is one that holds multiple meanings and unfortunately, has become ill-defined to those outside of the field. 
Thus, a second contribution of this paper is the broad scope of discussions around network analysis. In addition to discussions around social selection and social influence that are perhaps more familiar to network analysts, we also include topics that are of particular interest to psychologists, sociologists, and education researchers; these include network interference, co-evolution, ego network analysis, and network psychometrics. Thus, an additional goal of our paper is to provide guidance and layout potential pitfalls, underscoring what different network methods can and cannot achieve.


%
The remainder of the paper is organized as follows: we first introduce social network terminology and discuss exploratory methods. We then present models for social selection, discuss social influence, and methods for modeling co-evolution. We offer brief descriptions of ego-network analysis and network interference. Finally, we discuss network psychometrics and its similarities to social network analysis.

\section{Social Network Exploratory Methods}
We define a social network as the relationships among some set of individuals. The individuals in the network are called nodes, although nodes could also be entities such as schools or countries. The relationship is called a tie, which can be a connection such as friendship, coauthorship, or trade;  it could also be any connection linking two nodes including email correspondence, serving on the same committee, or liking someone's social media post. 

We collect social network data in a number of ways, from downloading large batches of online interaction data to surveying employees in a workplace and asking them to list their friends or favorite co-workers. These data are often represented as an $n \times n$ adjacency matrix $A$ such that the rows indicate the survey respondent and the columns indicate the person nominated. Thus, $A_{ij}=1$ would indicate that person $i$ listed person $j$ as their friend, and $A_{ij}=0$ otherwise. However, relationships are complex so we might be interested in something other than the presence and absence of a tie. For example, $A_{ij}=c$ indicates that $i$ retweeted $j$ $c$ times or that $i$ listed person $j$ as their friend and rated them a $c$ for closeness. We refer to these networks as having binary ties or valued ties respectively.

For simplicity, consider the following network data of seven people, and we will assume that these seven people listed their favorite coworkers. Thus, our toy network is binary and represents the absence or presence of a "favorite coworker" tie.  Using the data shown below, we see that node $2$ nominated node $1$ as one of their favorite coworkers and vice versa. Notice that node $5$ nominated node $1$ but node $1$ did not nominate node $5$. 
\begin{equation*}
\begin{array}{c|ccccccc}
& 1&2&3&4&5&6&7\\
\hline
1&0 & 1 & 1 & 0 & 0 & 1 & 0\\
2& 1 & 0 & 0 & 0 & 1 & 1 & 0\\
3& 1 & 0 & 0 & 1 & 0 & 1 & 0\\
4& 0 & 0 & 1 & 0 & 0 & 1 & 0\\
5& 1 & 0 & 0 & 0 & 0 & 0 & 0\\
6& 0 & 0 & 1 & 1 & 0 & 0 & 0\\
7& 0 & 1 & 0 & 0 & 0 & 0 & 0
\end{array}
\end{equation*}

 Because nominations for a favorite coworker need not be reciprocated, they result in a directed network and asymmetric adjacency matrix. Had we instead collected data on co-authorship, we would have had a symmetric matrix, also called an undirected network. 

\begin{figure}
    \centering
    \includegraphics[width=.5\textwidth]{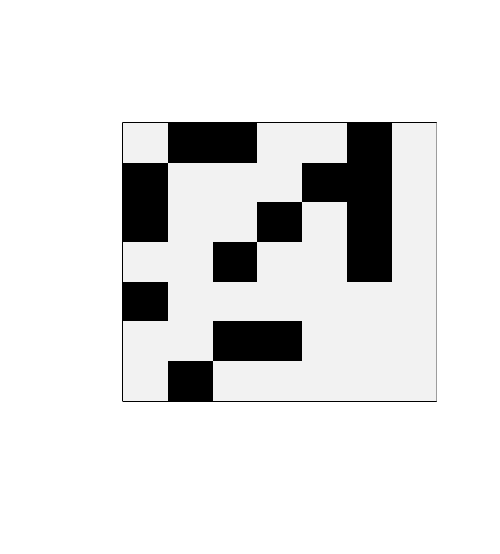}
    \caption{A visualization of the adjacency matrix; ties that are present are dark squares and absent ties are the light gray squares.}
    \label{fig:adjmatrix}
\end{figure}

Even when the network size (number of nodes) is fairly small, it can be difficult to capture the structure of the network from the adjacency matrix alone. These matrices are sometimes visualized using a grid plot in which boxes are colored to indicate the value or presence of a tie (see Figure \ref{fig:adjmatrix}). Grid plots can be useful for visualizing the overall saturation of the network---how much connection is visible in the network and how many nodes are not connected at all. Unfortunately because the resulting plot depends on the order of the nodes, network structure may not be visible without informed resorting of the nodes. Adjacency matrix visualizations are most helpful when group membership is known a priori and can be used to sort the nodes by group in the adjacency matrix.

\begin{figure}[hbt!]
    \centering
    \includegraphics[width=.7\textwidth]{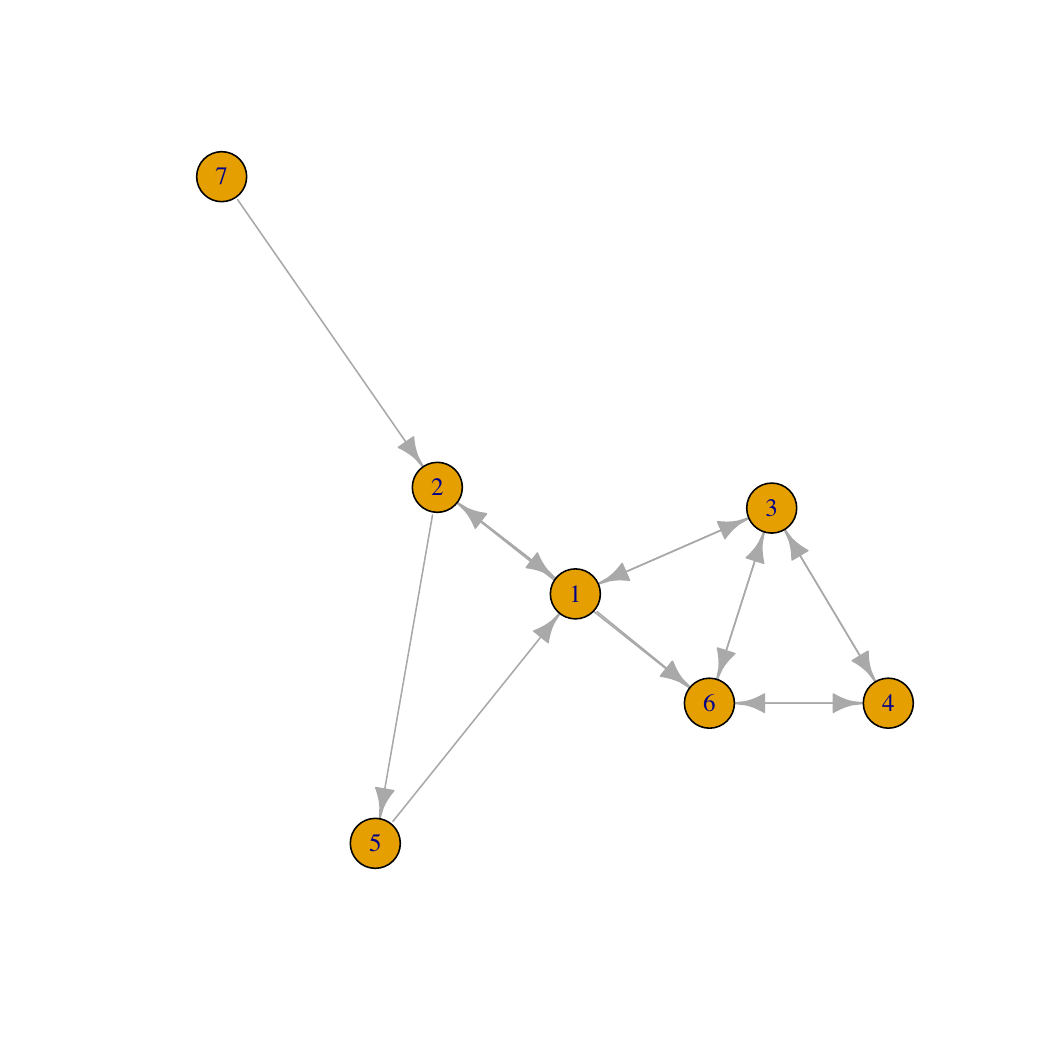}
    \caption{An example of a network with 7 individuals; the nodes are indicated by the vertices an the directed relationships are indicated by the arrows.}
    \label{fig:toyplot}
\end{figure}

When the network lacks known group structure or when the network has fewer than a few hundred nodes, the preferred way to visualize network data is with a network plot. In these plots, the nodes are plotted as vertices connected by edges based on the presence or the values of ties. Our favorite coworker network is shown in Figure \ref{fig:toyplot}. The arrows point from the employee who responded to the survey to the person nominated. Network plots can be helpful for visualizing the structure of the network as well; for example, nodes $3$, $4$, and $6$ have all nominated each other as favorite coworkers, and they appear together in the network plot.\footnote{The exactly positioning of nodes on a network plot is based on the plotting algorithm. Most software allows users to select an algorithm that positions nodes with ties near one another on the network plot.}  

\subsection{Descriptive Statistics}
There are a number of descriptive statistics used to quantify various aspects of a social network. We categorize network statistics into three classes based on what aspect they summarize: the network, the nodes, or the edges. Network statistics produce a single quantity to describe features of the entire network. As mentioned earlier, visualizations of the adjacency matrix are useful for gauging the overall saturation of the network. Density is a statistic that quantifies how connected a network is; it is defined as the average value across all possible ties. For a binary network, this is the proportion of observed ties and for a valued network, it is the average tie value. Networks with low density are called sparse, and networks with high density are called dense or saturated.

Transitivity, also called the clustering coefficient, is a term that measures how likely we are to see a clique or subgroup among three nodes. It is often calculated as the rate of observing two ties among three nodes such that a third tie is also observed. In other words, if nodes $i$ and $j$ share a tie and nodes $j$ and $k$ share a tie, we are interested in the proportion of times that nodes $i$ and $k$ also share a tie, resulting in completed triangle. This measure is related to clustering because networks with highly dense subgroup clusters - those that share ties within their clique and not between cliques -  will have many completed triangles.

The third common network statistic is called reciprocity and is calculated only in directed networks. Reciprocity refers to the proportion of reciprocal ties, and is usually calculated as the proportion of ties from $i$ to $j$ that also result in a tie from $j$ to $i$. Recall in Figure \ref{fig:toyplot}, node 5 nominated node 1, but their tie was not reciprocated.

Density, reciprocity, and transitivity are important network statistics because they often vary across networks. Certain types of relationships result in denser or sparser networks, with more or less transitivity, and with more or less reciprocity. For example, a co-authorship network is more dense in the medical sciences than in quantitative methods. Friendship tends to be more reciprocal than advice-seeking, but the rates of reciprocity for both types of networks depend on the individuals in the network. Friendship ties among adolescents tends to be more transitive than email ties among work colleagues. 

Edge statistics reflect different ways to move from one node to another along network ties. For example, the path from node 7 to node 3 in Figure \ref{fig:toyplot} could go from 7 to 2 to 5 to 1 to 6 to 3, but the shortest path is from 7 to 2 to 1 to 3. The number of edges along the shortest path between two nodes is called geodesic distance. The most common edge statistic based on geodesic distance is called edge betweenness, the proportion of all shortest paths between all pairs of nodes that include that edge. Edge betweenness is important for understanding how information flows through a network and which pathways are vital.

Node statistics are based on summary information about each node in the network. A common one is the number of or total value of the ties connected to specific node; this is called degree. For directed networks, degree can be further partitioned into in-degree (ties pointing to a node or that node being nominated by others) or out-degree (ties pointing out of a node or that node nominating others). Nodes with high degree can be thought of a gregarious, popular, powerful, or influential, depending on the relationship defined by the network tie. Nodes with high degree are often considered central whereas nodes with low or 0 degree thought of as isolated or marginalized in the network. Other measures of node centrality include node betweenness and closeness. Node betweenness, like edge betweenness, is based on geodesic distance among all pairs of nodes. A node's betweenness is the proportion of shortest paths that include that node. Closeness is also based on the geodesic distance. The closeness for node $i$ is the inverse of the sum geodesic distances from node $i$ to all the other nodes in the network.

One example of using node statistics in psychology research is a bullying intervention study \citep{paluck2016changing}. Rather than randomly assigning students to participate in the intervention, the researchers chose students with high degree, whom they called social referents. These students were selected because of their high degrees of connections to others, increasing the likelihood of diffusion of the intervention.

\subsection{Community Detection}
Networks with subgroup structure are often of interest to researchers because understanding which nodes are in the same subgroup or cluster helps researchers theorize about network formation or persistence. Thus, community detection methods for identifying these subgroups or communities are a standard part of network exploration. 

\begin{figure}
   \centering
   \includegraphics[width=0.5\textwidth]{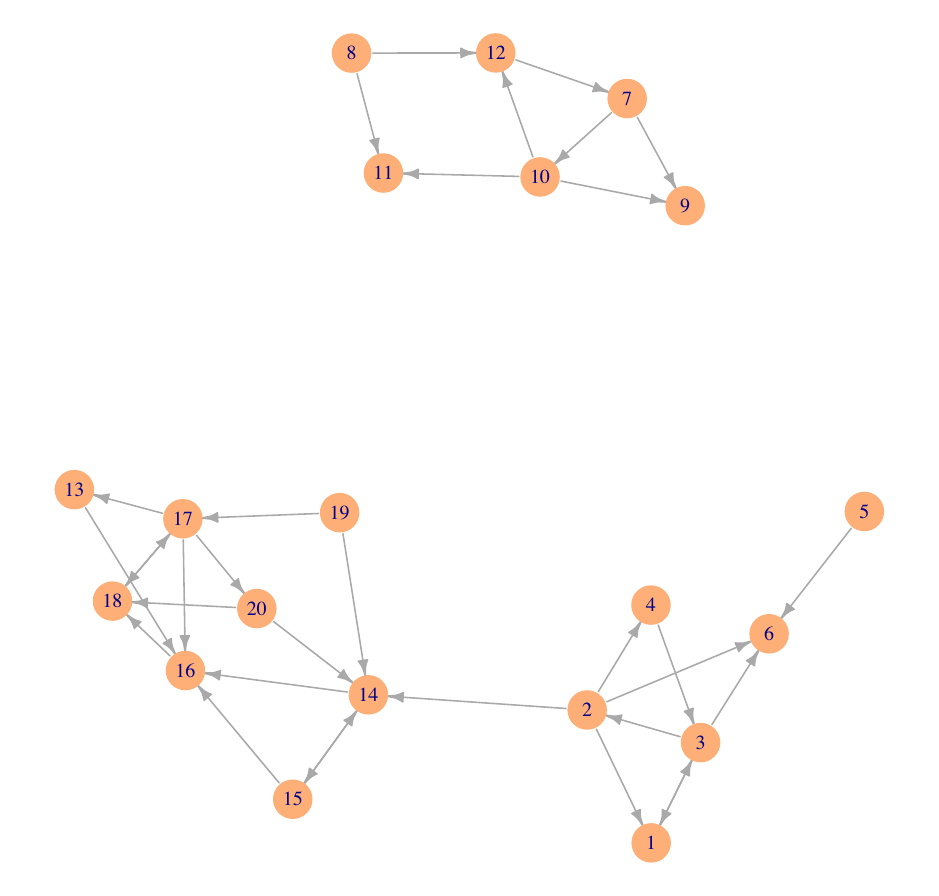}
   \caption{Example of network with subgroup structure; nodes are colored by subgroup membership}
    \label{community}
\end{figure}

We define a network community or subgroup as one that shares ties within the community at a higher rate than with those outside their community. As a result, the within subgroup density is generally higher than the between subgroup density, as shown in Figure \ref{community}. Based on this defining feature of differing densities, different community detection algorithms identify communities in different ways. There are number of algorithms, but we discuss the three most common algorithms.

The most intuitive algorithm is the random walk algorithm \citep{pons2006computing}. Random walks are exactly as they sound, we start at a random node and based on the ties coming from that node, we randomly follow an edge to one of the connected nodes at a random. For example, in Figure \ref{community}, if we start at node 2, we can walk to nodes 4,1,6, or 14, but not 3 because of the direction of the tie. From 3, we can walk to 2, 1, or 6. The random walk repeats for $k$ steps, and the nodes that are included in the walk are saved. This process is repeated many, many times, and communities are based on the nodes that tend to be in the same random walk. This intuitively makes sense because if two nodes are in the same community, they will likely be part of random walks involving those nodes.

Two other algorithms are called edge betweenness \citep{girvan2002community} and fast greedy \citep{clauset2004finding}. We will discuss them together since they are inversely similar. The edge betweenness algorithm is divisive in that all nodes begin in the same community and are iteratively separated. At each iteration, the edge with the highest edge betweenness is removed, eventually splitting the nodes into $g+1$ communities. This algorithm continues until all nodes are in separate communities. Fast greedy is an agglomerative algorithm that starts with all nodes in separate communities and iteratively merges two communities until all nodes are in the same community. The solution for both of these algorithms, as in where along the algorithm we should stop, is usually determined by modularity. This is a measure based on the observed edges within each community and how many edges would be expected if the edges were randomly distributed \citep{newman2004finding}.

Community detection and modularity has been used in mapping regions of the brain \citep{stevens2012functional} and mapping language networks \citep{siew2013community}. 
Further, understanding subgroup structure and community membership is important for many psychological disciplines, particularly in studies of relationships, inclusion/exclusion, peer effects, and information diffusion. First, understanding the extent to which a network has subgroup structure is important for social in-group or out-group studies. Peer networks in which there are very distinct social groups operate differently than networks that are highly integrated. There may not be inter-group exclusion if distinct groups are absent from the network. Similarly, studies on information diffusion such as misinformation or gossip will manifest differently in networks with highly insular subgroups. 



Finally, understanding community structure and membership is important for designing experiments aimed at changing behaviors and influencing outcomes when individuals are part of a larger social system, such as a workplace or school. For example, \cite{paluck2016changing} selected social referents in schools to participate in bullying intervention.  The researchers could have also conducted community detection on these networks; if there were distinct social groups, the researchers could have selected the social referents from separate social groups to participate in the intervention, thereby further increasing the likelihood of diffusion of anti-bullying sentiment.


Network statistics and community detection are examples of exploratory or descriptive methods. The remainder of this manuscript focuses on social network models, which are methods for inference. Akin to regression models that link independent and dependent variables, social network models allow us to quantify associations between networks and covariates. Therefore, networks can operate as either independent variables or dependent variables. With social network analysis, we can investigate how a network - or the people with whom we interact - impacts our outcomes; here the network is the independent variable. We can also investigate why a network forms or why two individuals interact, where the network is the dependent variable. In the following sections, we consider inferential models that directly address the relationships between networks and behaviors/outcomes.

\section{Social Selection} 

Social selection describes the process of forming relationships as people connect based on similar interests, geographic locations and joint activities \citep{antonoplis2022has,verdecho2011review}. Social selection models are used to answer questions how relationships are formed and why we form relationships with some but not others. Social homophily is one mechanism for  forming connections and is based on shared interests, beliefs and behaviors. On the other hand, we also form relationships based on dissimilarities as we seek advice from those who have insights we do not yet possess or perspectives we do not have. \\

 Social selection research has the potential to address urgent societal concerns and provide meaningful insights into the formation of the current political and social climate. 
Racially segregated social groups has been shown to potentially perpetuate unequal access to information and opportunities, exacerbate wealth inequality and create unhealthy attitudes towards the out-group \citep{skinner2023systemic}. Social selection models could be used to investigate the underlying motivations of racial homophily, whether stereotypes contribute to racial segregation and discriminatory behaviors, which and to what extent individual choices, attitudes and behaviors can explain racial segregation. Existing research has also highlighted the potential benefit of racial homophily for members of marginalized groups, leading to greater relational intimacy and self-esteem \citep{mcgill2012intra}. 

Consider the study by \cite{antonoplis2022has}, in which they found a positive association between a measure of openness to others and cross-racial friendships. Across multiple studies, they asked individuals to list friends and then categorize their race; the proportion of individuals of a different race was positively associated with openness to others. While the study discussed social networks, a sociometric network was not collected. It is possible, therefore, that those individuals who rated themselves high on openness to others would also find having cross-racial friendships as socially desirable and would be more motivated to inflate their cross-racial friendship counts. A sociometric network approach would allow researchers to formally define friendship as individuals mutually nominating each other as friends. Further, there are additional research questions that could be answered with social selection models. We could explore to what extent certain combinations of racial pairings are more common; and to what extent friendship ties are reciprocated based on race and openness to others.


\subsection{Social Selection Models}
In social selection models, the networks are the responses, and individual attributes and behaviors are incorporated as predictors. In this way, social selection models allow researchers to relate attributes of the nodes to their relationships and answer questions related to how relationships are formed.
One notable thing is that network ties are dependent observations. Person A's connection with person B is related to person A's connection with person C as both connections are impacted by the friendship seeking behavior of person A and whether A is actively seeking connections; simultaneously, person B's connection with person C is impacted by their shared connection with person A, and perhaps the two have interacted through their mutual connection with A. Unlike other data situations where observations that are easily assumed to be independent, the interconnectedness of relationships is a key network feature as well as a major hurdle for network modeling, and standard statistical models are not appropriate.

In addition to modeling social relationships, social selection models have also been applied to understand and accommodate relational networks in other contexts including conflict events \citep{dorff2023network}, international trade relations \citep{minhas2016new}, company board affiliations \citep{friel2016interlocking}, terrorist recruitment \citep{wang2022resilience}, structural and functional brain connectivity \citep{wang2019symmetric,wang2021learning,liu2021graph} and other biological networks \citep{gollini2016joint}. 



Thus, the remainder of this section provides  an overview of common social selection methods for modeling social network data. In particular, we organize our presentation of network models by how these methods address dependencies among network ties, and we define two large categories of models based on common usage. We introduce exponential random graph models \citep[ERGMs,][]{wasserman1996logit} and latent variable network models, which include stochastic block models \citep[SBMs,][]{holland1983stochastic,airoldi2008mixed} and latent space models \citep[LSMs,][]{hoff2002latent}. Between ERGMs and latent variables-based methods, the difference lies on whether we explicitly or implicitly account for dependencies among the connections in the network. ERGMs assume that dependencies among the connections can be accounted for by descriptive features of the network. In latent variables-based methods, we condition the dependencies out of the model using latent variables. These methods are also called the conditionally independent dyad (CID) network models \citep{CIDpaper} because the models assume conditional independence among network ties conditional on the parameters and latent variables in the model.




\subsubsection{Exponential Random Graph Models}
Exponential random graph models \citep[ERGMs,][]{wasserman1996logit, robins2007introduction} define the probability of observing the network $A$ as 
 \begin{align}
     p(A_{ij} | \boldsymbol{\theta}) = \frac{\exp(\boldsymbol{\theta}^T h(A))}{c(\boldsymbol{\theta}, A)}, \label{ergm}
 \end{align} where $h(A)$ represents a collection of statistics describing the features of the network, parameterized by $\boldsymbol{\theta}$; and we use $c(\boldsymbol{\theta}, A)$ to represent the normalizing constant. The Erdos–Renyi–Gilbert model \citep{gilbert1959random}, for example, accounts for the total number of ties in the network. Other substructures of interest include reciprocity, the degree of each node, the number of k-cycles, and the number of k-paths (see \cite{wasserman1994social} for additional examples). Each formulation generates a probability distribution for one possible network with $n$ nodes out of a total of $2^{\frac{n(n-1)}{2}}$ possibilities. 

Because the explicit dependence structure is uniquely defined by the network statistics included in the model, simple dependence structures were originally proposed. For example, Markov dependence assumes dependence between dyads and among ties between adjacent nodes \citep{FrankStrauss1986}. In practice however, using network statistics that map to a clearly defined dependence structure can result in model degeneracy, predicting completely dense or completely sparse networks \citep{handcock2003assessing}. As a result, it is now common practice to use geometrically weighted degree distributions or similar geometrically weighted statistics in an ERGM \citep{Snijders2006ergms}.

In the past two decades, there has been a plethora of research on ERGMs. A number of estimation strategies has been proposed for extremely large \citep{stivala2020exponential} or small networks \citep{stivala2020exponential}. In addition to work on estimation, a number of extensions to standard ERGMs have also been proposed. Readers interested in longitudinal versions of ERGMs may be interested in work by \cite{robins2001network} and \cite{HannekeFu2010}; for multilevel or multiplex versions of ERGMs, we recommend \cite{slaughter2016multilevel}, \cite{lazega2015multilevel}, and \cite{wang2013exponential}.

\subsubsection{Latent Variable Network Models}
Although ERGMs are popular modeling approaches, we recommend latent variable network modeling approaches for several reasons. First, many psychology researchers are very familiar with latent variable models, and latent variable model courses on SEM or factor analysis are common in graduate school. In addition, social network latent variable approaches are more intuitive for understanding how the data came to be, and these models are generally easier to estimate without degeneracy issues. 

Latent variable network models accommodate inter-tie dependence implicitly through latent variables. Thus, conditional on the parameters and latent variables in the model, network ties are assumed to be independent. This assumption allows these models to be written as data generating models, and estimation is more straight-forward \citep{CIDpaper}. We will introduce two classes of latent variable network models: stochastic block models \citep[SBMs,][]{holland1983stochastic} and latent space models \citep[LSMs,][]{hoff2002latent}. 

\paragraph{Stochastic Block Models}
Stochastic block models assume that connections are independent conditional on latent group membership and the probability of a tie is the same for any node in the same group. Specifically, we define a stochastic block model in terms of the probability of observing network $\boldsymbol{A}$ given latent group membership $\boldsymbol{U}$ as the product of observing each tie $a_{ij}$ conditional on latent group membership $u$, which is given as
 as \begin{equation}
    P(\boldsymbol{A}|\boldsymbol{U}) = \prod_{i=1}^N \prod_{j=1, j \neq i}^N p(a_{i,j}| u_i, u_j), \qquad p(a_{i,j}| u_i, u_j) = m_{u_i,u_j}, \label{sbm}
\end{equation}where $u_i$ is a discrete latent variable representing the group membership for node $i$, $u_i \in {1,2,...,K}$, where $K$ is the total number of blocks or groups; and $m$ is a $K \times K$ block probability matrix such that $m_{u_i,u_j}$ represents the probability of a tie from the group that $i$ belongs to the group $j$ belong. In a simple block model in Equation \eqref{sbm}, the probability of two nodes forming an edge is independent of all the other edges in the network conditional on the nodes' block memberships. 

We often assume that $m_{u_i,u_i} > m_{u_i,u_j}, i\neq j$ so that nodes in the same block are more likely to have ties than nodes in different blocks. Under this assumption, one can see how this model would generate networks with cluster structures.

Regardless of the values of $m$, SBMs produce networks where members of the same cluster show more similar patterns of connections with each other than members of different clusters. Perhaps members of the cluster A tend to be strongly connected with members of cluster B, but not members of cluster C. Such network patterns can be captured by SBMs, which identify and assign structurally equivalent nodes in the same cluster based on their relationship pattern across the whole network. 


\paragraph{Latent Space Models}

While SBMs generally requires fit data with some kind of block or subgroup structure, latent space models \citep[LSMs,][]{hoff2002latent} and their extensions accommodate any type of network. Latent space models assume that each node in the network has a position in a latent social space; network ties are independent conditional on these positions and any other parameters in the model.The probability of a tie is a function of the latent positions and given as
\begin{equation}
 P (\boldsymbol{A}| \boldsymbol{U}) = \prod_{i=1}^N \prod_{j=1, j \neq i}^N p(a_{i,j}| \boldsymbol{u}_i,  \boldsymbol{u}_j)), \qquad p(a_{i,j}|  \boldsymbol{u}_i,  \boldsymbol{u}_j) = g( d(\boldsymbol{u}_i,  \boldsymbol{u}_j)),  \label{lsm}
\end{equation}where $\boldsymbol{u}_i$ is a continuous latent variable describing the latent position for node $i$; $\boldsymbol{u}_i$ is usually multidimensional with a low dimension of 2 or 3 \citep{hoff2002latent}. The value of $a_{ij}$ is thus defined to be a function of some distance metric $d()$ between $\boldsymbol{u}_i$ and $\boldsymbol{u}_j$. The function $g()$ is a link function to accommodate the value of the network tie analogous to link functions in generalized linear models. For binary network ties, the logistic link is often used; for ordinal ties, a probit link can be used; and for continuous tie values, the identity link can be used.


The other specification needed is the distance metric $d()$. We typically choose a distance such that closer nodes are more likely to be connected (or more likely to have a higher tie value in a continuous tie network) than nodes that are far apart. When \cite{hoff2002latent} originally proposed latent space models, $d()$ was defined as the Euclidean distance such that $d(\boldsymbol{u}_i, \boldsymbol{u}_j) = \vert \boldsymbol{u}_i -\boldsymbol{u}_j \vert$. Another commonly used distance metric is the inner product \citep{hoff2005bilinear}. This metric is given is $d(\boldsymbol{u}_i, \boldsymbol{u}_j)=\boldsymbol{u}_i^T \boldsymbol{u}_j$, and the model was termed the bilinear mixed effects model when random effects were also incorporated. While distance metrics by definition are commutative, latent space models can also accommodate directed network data by incorporating random effects, such as sender and receiver effects, as well as covariates. 

In recent years, the additive and multiplicative effects model  \citep[AME,][]{minhas2019inferential,hoff2021additive} introduced another latent effect defined by $d(\boldsymbol{u}_i, \boldsymbol{v}_j)=\boldsymbol{u}_i^T \Lambda \boldsymbol{u}_j$, where $\Lambda$ is a diagonal matrix. Note that this combination of latent variables does not produce a symmetric distance, therefore allowing the latent vectors to have different sender and receiver contributions. With multiple latent variable configurations, random effects, and covariates, latent space models allow researchers to answer substantive questions about social selection, positioning latent space models as ideal models for social selection. 

Finally, there has been a surge of development in the area of latent variable network models, and we briefly give some examples. These models have been adapted to accommodate multi-layer networks \citep{gollini2016joint,salter2017latent,arroyo2019inference}, multilevel networks \citep{sweet2013jebs}, dynamic networks \citep{sewell2015latent,friel2016interlocking,sarkar2007latent} and others \citep{wang2023joint,gu2022joint}.

\subsection{A Social Selection Example in Psychology}
Let us return to the study by \cite{antonoplis2022has} about openness to others and interracial friendships. If sociometric social network data had been collected, the authors could have used a latent space model to determine the association between openness and interracial friendship. Define $A_{ij}$ as node $i$ nominating node $j$ as a friend where node $i$ and $j$ identify as different races. By including a covariate about openness $X$, we could fit the latent space model given as 

\begin{equation}
    \mbox{logit}(P(A_{ij}=1) = \beta_0 + \beta_1 X_i + \beta_2 X_j - \vert u_i - u_j \vert~,
\end{equation}where $u_i$ is the low dimensional latent space position for node $i$. We include two covariate effects such that $\beta_1$ describes the effect of openness on nominating a person of a different race as a friend and $\beta_2$ describes the effect of openness on being nominated by someone of a different race as a friend.

Not only could we have examined differences in sending and receiving interracial tie nominations, social network data also allows us to answer other research questions such as which types of interracial friendships are most likely and to what extent does homophily impact interracial friendship. Unfortunately, we do not have social network data to address openness and cross racial friendship, but we present in this section an example rooted in the context outlined by \cite{antonoplis2022has}. 

\begin{figure}[hbt!]
\centering
  \includegraphics[scale=.43]{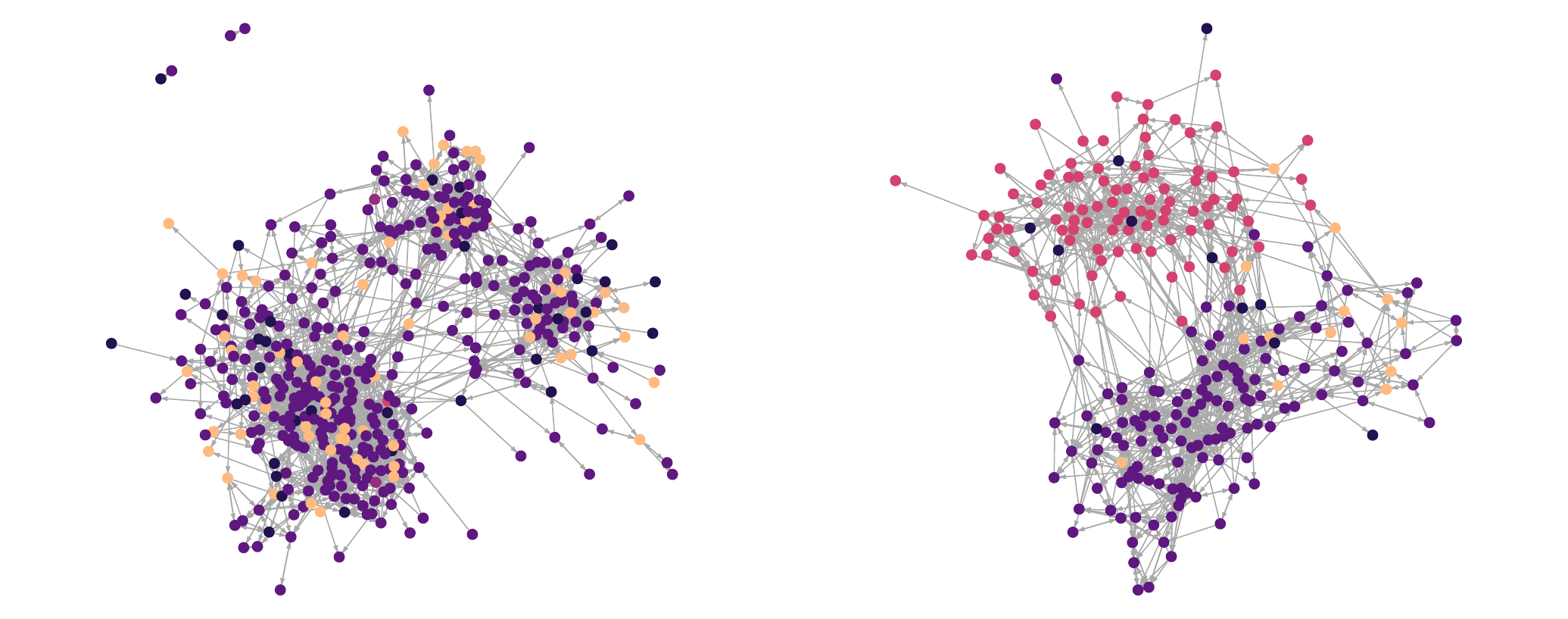}
 \caption{A. The network plots of two schools in the National Longitudinal Study of Adolescent to Adult Health (Add Health) study. The nodes represent students, colored by race (black, purple, pink, and  orange represent race coded as Hispanic, White, Black, and Other respectively), and the edges represent friendship ties.}
\label{sse}
\end{figure}

Racial homophily in the US is a likely phenomenon due to the historical contexts of slavery, white supremacy, and anti-Black racism, which has largely contributed to people's perceptions of those from differential racial groups \citep{antonoplis2022has} and segregation of friendship groups. Thus, we explore racial homophily through the lens of social selection using two student school communities collected in the National Longitudinal Study of Adolescent to Adult Health \citep[Add Health;][]{harris2019cohort}. We fit a latent space model model with student friendship as the response. The networks of the two schools are shows in Figure \ref{sse} where the nodes are colored by race. In the first school (left), 73\% of the students are white and the next largest group is Hispanic students who makeup only 8.5\% of the school. There does not appear to be strong racial homophily compared to the second school (right) where self-clustering of students by race is apparent. In this school,  52\% of the students are white and 38\% of the students are Black. 

We fit separate latent space models to these schools but include two dyad-level covariates. We include an indicator that students are the same race and an indicator that if the students are not the same race, they are both non-white students. We make the following inferences based on the 95\% equal-tailed credible intervals shown in Table \ref{tab:my_label}. In School 1, we do not see evidence of racial homophily although students of different minoritized racial groups are less likely to be friends. In School 2, there is a positive effect of racial homophily and no evidence that minoritized students of different races are more or less likely to be friends. 

\begin{table}[]
    \centering
    \begin{tabular}{c|c|c}
    \hline
    & School 1 Mean (95\% CI) & School 2 Mean (95\% CI)\\
    \hline
         Same Race& 0.00 (-0.07, 0.08)  & 0.87 (0.98, 1.10) \\
         Diff Race, non-white& -0.14 (-0.24, -0.05) & 0.02 (-0.11,0.17) \\
         \hline
         \end{tabular}
    \caption{Social selection latent space model fits for School 7 and School 9 for the association between racial homophily and friendship ties}
    \label{tab:my_label}
\end{table}

\section{Social Influence}
Social influence is the process of individuals changing their beliefs, attitudes, or behaviors based on the beliefs, attitudes or behaviors of people in their networks. Research on social influence or peer effects is relatively common in some disciplines such as developmental psychology, social psychology, educational psychology, and organizational psychology. For example, \cite{hamamoto2021examining} found evidence of social influence in how children arranged stickers when they were working with adults, and \cite{leung2022susceptibility} explored the extent of social influence susceptibility following mindfulness training. In education, \cite{fleming2001benefits} and \cite{saleh2007} examined the impacts of collaborative learning and found that some of the positive outcomes of those studied are due to behaviors gained through peer social influence. In addition, there is also evidence that relationships with peers are predictive of academic outcomes \citep{white2021peer, risi2003children}. 
Among adults, social influence is prevalent in the organizational psychology literature. \cite{carnevale2022laughing} found that the type of a leader's style of humor impacted workers' outcomes, and \cite{leblanc2022leader} found that under certain circumstances, leader humility increased team reflexivity. While there is interest in social influence research, the use of social influence models to quantify the extent to which individuals are influenced by others is rare. That is, researchers in psychology often assume peer effects or influence, and use a positive association between individual outcomes and their peers' outcomes as evidence of social influence. We believe that social influence models could be used to more formally test or measure the extent to which this assumption is true. From the literature, \cite{molloy2011peer} stands out as a paper that used social influence models to examine the effects of different types of network relationships on academic motivation.

For this paper, we consider the study by \cite{KIM2014134}, which explored the mediating effects of peer and instructor enthusiasm on the relation between initial subject matter interest and measures of subject matter interest measured at a later time point. Their study included over 400 undergraduates taking courses in 12 different disciplines. They found that perceived peer enthusiasm and perceived instructor enthusiasm predicted both measures of situational interest and further found that enthusiasm mediated the relationship between initial interest and end of course interest. In \cite{KIM2014134}, the researchers assumed the presence of social influence and used initial interest to predict future interest with enthusiasm as a mediator. Had these researchers been interested in the effect of peer interest on course interest, they could have collected social network data and fit a social influence model. Given the scale of this study, a sociometric analysis would have been difficult, but an ego network analysis to measure social influence could have been done. For example, researchers could have included items in the survey that asked about the levels of interest of the peers with whom they spent the most time or to whom they felt closest. Subsequent fitting of a social influence model would then quantify the extent to which social influence impacted interest in the course.

\subsection{Social Influence Models}
Whereas social selection models require specialized models to account for the ill-defined interdependence among network ties, general and generalized linear models can be used as social influence models. The intricacies instead lie in the assumptions of the data.

For example, a model for social influence can be written as 
\begin{align}\label{e:influence-generic}
    Y= \beta X + \theta g(A, Y*) + \epsilon~,
\end{align}
where $Y$ is a node-level outcome such as a behavior or belief, $X$ is a collection of covariates, $A$ is the social network, and $g$ is a function of that network and a version of the outcome $Y*$. The parameter $\theta$ quantifies the amount of social influence in the network.

In fact, it is more common to augment Equation \ref{e:influence-generic} to include earlier measurements of $Y$. That is, $Y*$ and $X$ both include an earlier measurement of $Y$.  This model is given as
\begin{align}\label{e:influence-temporal}
    Y^t= \beta_0 + \beta_1 Y^{t-1} + \theta g(A, Y^{t-1}) + \epsilon~,
\end{align}
such that someone's outcome $Y$ at time $t$ is a function of their outcome at an earlier time $t-1$ and their peers' outcomes at that earlier time. Having before and after measurements of the outcome $Y$ allows the network ties to impact changes in $Y$ as opposed to $Y$ being correlated with peer outcomes. Versions of this model have been called a diffusion model \citep{valente2005network} and a social influence model \citep{frank2014social}. 

Equation \ref{e:influence-temporal} includes a function $g$, and a common definition of $g$ is to create weights defined as $W_i=\frac{A_{ij}}{\sum_i A_{ij}}$ and then $g_i(A, Y^{t-1})=\langle W_i, Y^{t-1}\rangle$, where $g_i$ is the dot product of $W_i$ and $Y^{t-1}$. Thus, $g_i$ is the average of $Y^{t-1}_j$ for all $j$ that is sent a tie from $i$, if the network is directed. If the network is undirected, $g_i$ is the average of $Y^{t-1}_j$ for all $j$ that share a tie with $i$. 

Other definitions of $g$ are possible. For directed networks, we can specify whether influence should include only the nodes that are nominated by $i$ (as previously defined) or only nodes that nominate $i$ or both. We could further extend our definition of $A$ and instead use $A^2$ so that node $i$ is influenced not only by their own ties, but also the ties of their neighbors. Another possibility is to use a latent space model and extract relative influence based on the pairwise distance between nodes so that nodes are most influenced by those closest to them in the latent space \citep{Sweet2020Influence}. 

Returning to Equation \ref{e:influence-temporal}, having specified $g$, we can see that $\theta$ is the effect of social influence; that is, $\theta$ quantifies the relative impact of one's peers on the outcome of interest. 

Because social influence is a causal process, we must make several strong assumptions. First, we assume that the network $A$ existed between the time points $t-1$ and $t$. Having temporal precedence allows us to assume that influence occurred, rather than selection. If network $A$ is formed close in time to when $Y^t$ is measured, it is unclear if outcome $Y^t$ is the result of network $A$ or if rather $Y^t$ caused network $A$. Second, we must assume that we have controlled for all confounding variables in our model; that is, any variable that could have impacted $Y^t$ should be included. This is why having $Y^{t-1}$ in the model is important, but other considerations are needed to assume that Equation \ref{e:influence-temporal} is correctly capturing social influence. We will discuss these assumptions in a real-world example below. 

Other models of social influence exist, but they require even stronger assumptions. One example is the network autocorrelation model \citep{doreian1989network, leenders2002modeling}. This model is commonly written as

    \begin{align}\label{e:influence-autocorr}
    Y= \rho W Y + \beta X + \epsilon~,
\end{align}
where $Y$ is the outcome of interest, $X$ is a set of covariates, and $W$ is the network weight matrix. Then $\rho$ is defined as the autocorrelation parameter, and a hypothesis test for whether $\rho=0$ is used to establish social influence. 

Because this model is an autocorrelation model, we assume that the network $Y$ has achieved some level of stationarity; that is, any process that is impacting the network weights $W$ is completed and that $Y$ does not impact $W$. This assumption is quite strong, and we believe it is rarely met. For our purposes, we will focus on the temporal influence model given in Equation \ref{e:influence-temporal}.

\subsection{A Social Influence Example in Educational Psychology}\label{s:Influence-Example}
Our example is inspired by that of \cite{KIM2014134} that examined the effects of instructor and peer enthusiasm on student interest in their course. Suppose we instead want to quantify the amount of social influence that peers have in a university classroom. For simplicity, let us assume our outcome is STEM interest. We could collect STEM interest early in the semester and again at the end of the semester. During our end of semester survey, we would ask students to list those students in the class with whom they spent academic time during that semester. 

\begin{figure}[hbt!]
    \centering
    \includegraphics[height=65mm]{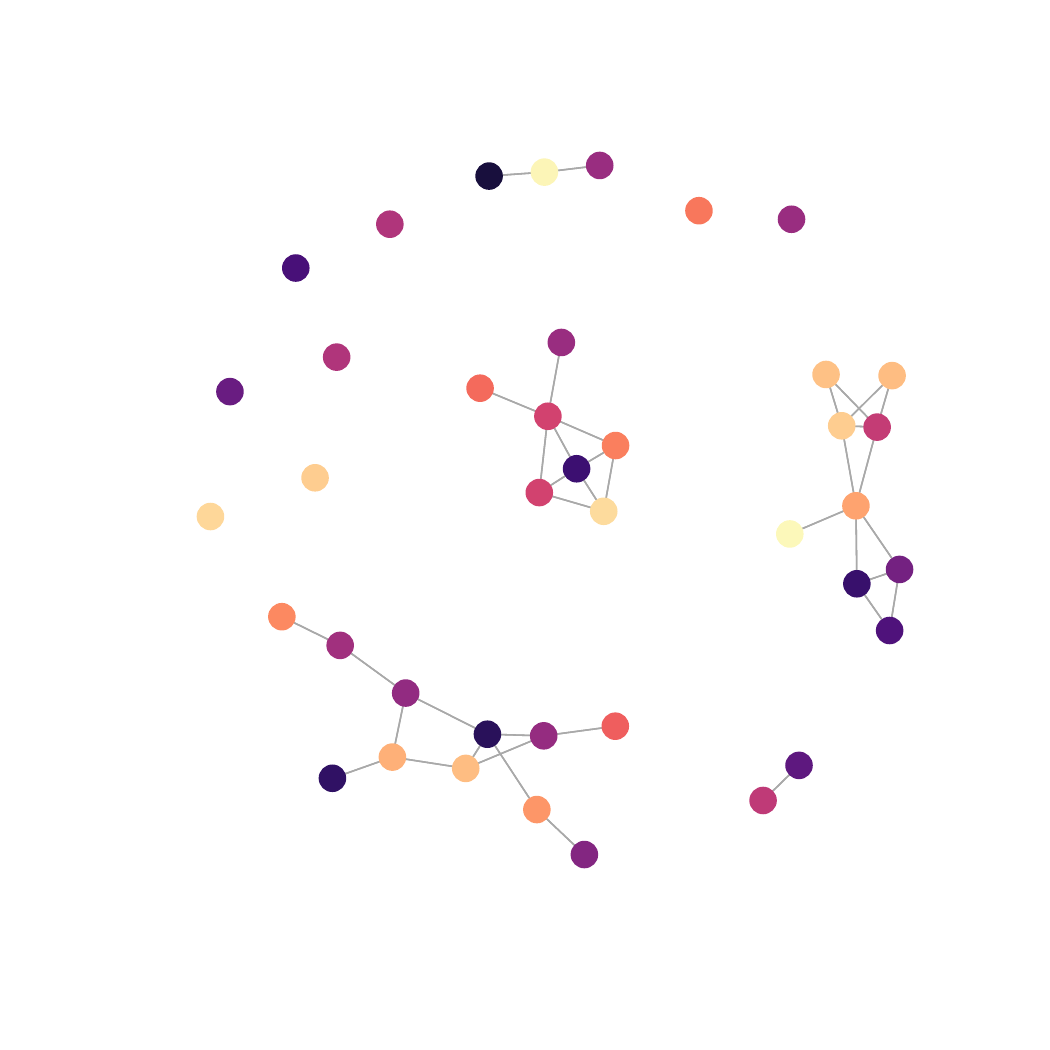}\hspace{1cm}   \includegraphics[height=65mm]{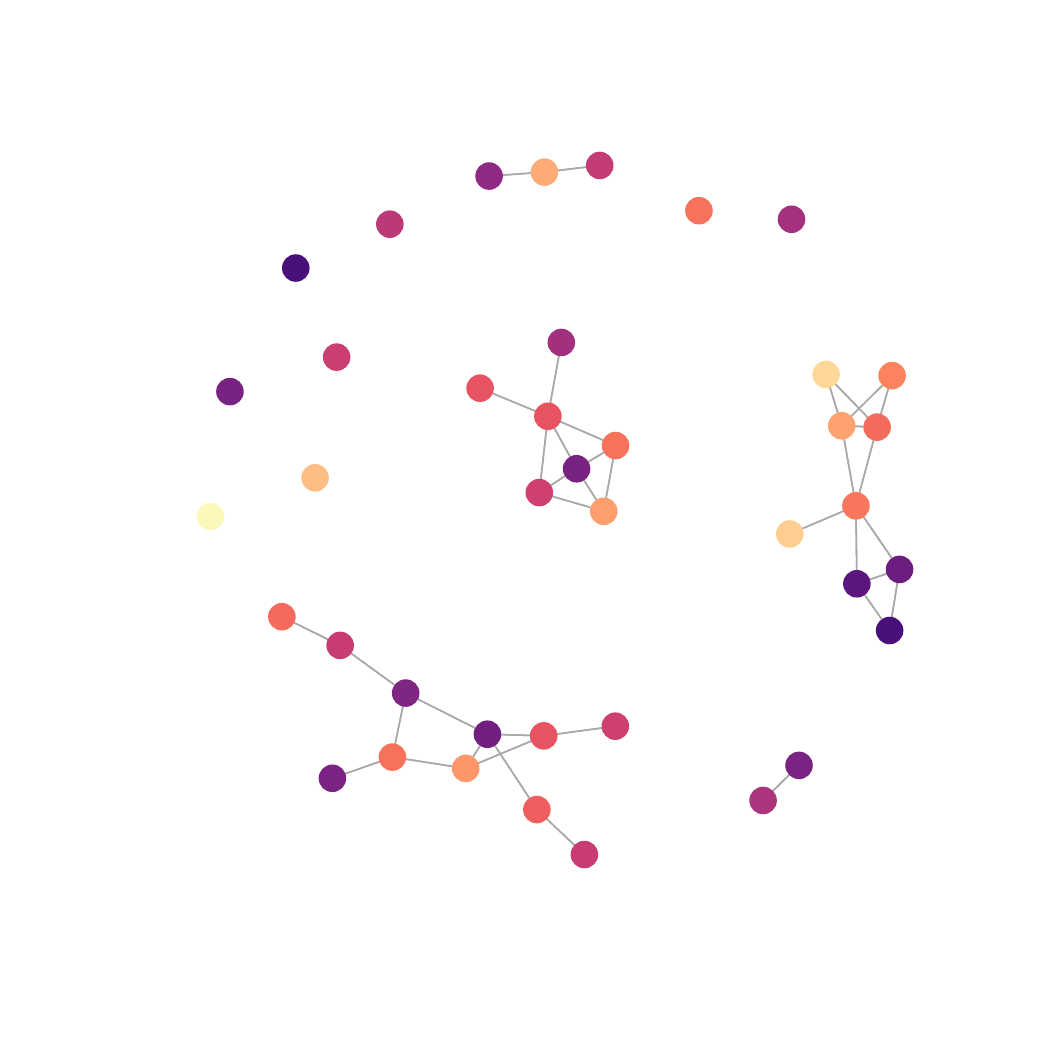}
    \caption{Hypothetical study network among college student shows network structure where nodes are colored by interest measured at the beginning(left) and end(right) of the course.}
    \label{fig:influence.example}
\end{figure}

From this scenario, we have data for $Y^t$, $Y^{t-1}$, and $A$, where $Y$ is course interest measured at time $t-1$ and time $t$. Figure \ref{fig:influence.example} (left) shows our hypothetical university class of students, who are connected if they interact at least once a week for this class. The nodes are colored (light beige to pink to magenta to dark purple) based on STEM interest at the beginning of the semester; we assume darker nodes indicate more interest in STEM.

Figure \ref{fig:influence.example}(right) shows how interest in STEM changed as a result of social influence at the end of the course. Students who did not interact with others changed very little during the course; any fluctuations in interest can be attributed to random variation. Students who did interact with each other during the course were subject to influence. Based on moderate influence findings from studies on peer reading \citep{cooc2017peer} and mathematics pedagogy \citep{spillanehopkinssweetBELIEFS}, we could assume a moderate social influence effect.

 

Using this toy example, we could fit the social influence model given in Equation \ref{e:influence-temporal}, where $g(A, Y^{t-1})_i$ is the average course interest of all the individuals with whom node $i$ regularly interacted. 


Our fitted model has the parameter estimates: $\hat{\beta_1} = 0.73$ and $\hat{\theta}=0.15$, both statistically significant at the 0.05 level. A positive value for $\theta$ indicates the presence of social influence in that individuals' STEM interest became more similar to the STEM interest of those with whom they interact. 

This example serves to illustrate how to quantify the extent to which social influence is present in a social system, but extensions could also be explored. For example, we could have examined how individual attributes could have moderated the amount of social influence in the network. 

Let us consider such a moderation model. Suppose we include a measure for self-esteem $X_{se}$ in our model given as:

\begin{align}\label{e:influence-temporal-cov}
    Y^t= \beta_0 + \beta_1 Y^{t-1} + \beta_2 X_{se} + \theta g(A, Y^{t-1}) + \epsilon~.
\end{align}

When fitting Equation \ref{e:influence-temporal-cov}, it is apparent that $\beta_2$ accounts for any differences in STEM interest at time $t$ for different self-esteem levels, conditional on STEM interest at time $t-1$ and peers' interest. But this model may not fully encapsulate our theory about self-esteem, STEM interest, and social influence. Perhaps we are interested in whether social influence on STEM interest differs by self-esteem; that is, are students with lower self-esteem more likely to become interested in STEM because of their friends? 

The necessary model includes an interaction term between the peers' STEM interest and self-esteem and is given as: 

\begin{align}\label{e:influence-temporal-mod}
    Y^t= \beta_0 + \beta_1 Y^{t-1} + \beta_2 X_{se} +  \theta g(A, Y^{t-1}) + \omega g(A, Y^{t-1})X_{se}+ \epsilon~.
\end{align}

\begin{figure} [hbt!]
    \centering
    \includegraphics[width=.9\textwidth]{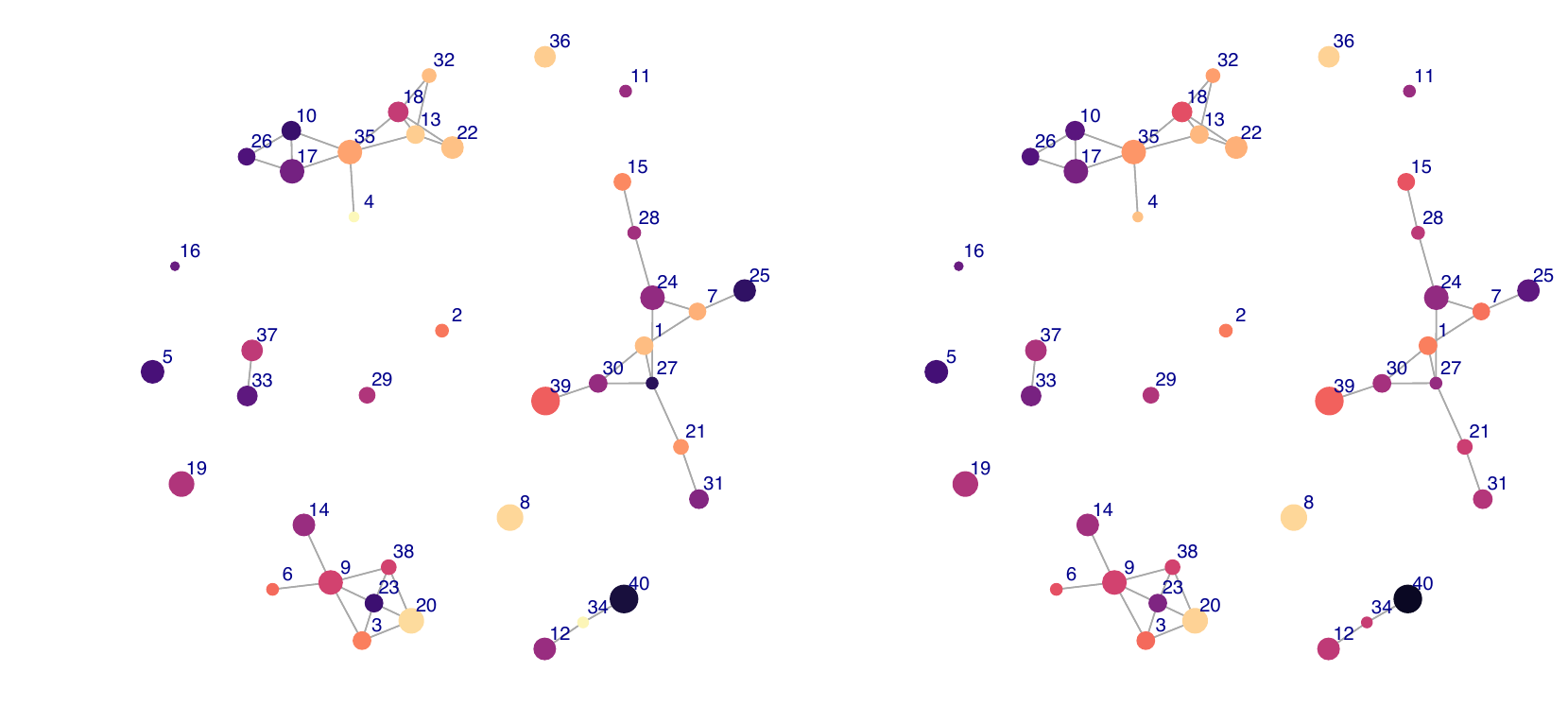}
    \caption{Example of a moderating effect of social influence where the smaller nodes are more susceptible to social influence than the larger nodes.}
    \label{fig:moderation}
\end{figure}

Figure \ref{fig:moderation} (left) shows an example of a friendship network at time $t-1$ where the nodes are colored based on STEM interest and node size maps to self-esteem. Suppose that the darker colors indicate more STEM interest, and larger nodes have higher self-esteem. If our hypothesis is correct, we would expect the larger nodes to be less susceptible to social influence than the smaller nodes. We used a simulated example to illustrate such an interaction effect between STEM interest and self-esteem in Figure \ref{fig:moderation} (right). In this figure,  students with high self-esteem, represented as large nodes such as 9, 39, and 40 do not change their STEM interests despite being connected to nodes with different STEM interest levels, while students with low self-esteem, small nodes such as 4, 27, and 34 tend to be more influenced by others.

\begin{table}[h]
\centering 
\caption{Model results for the interaction between social influence and self-esteem.}
\label{tab:social.inf}
\begin{tabular}{llr}
\hline
    Parameter & Estimate & SE  \\ 
    \hline
    $\beta_0 $ & -0.01 & 0.85\\
    $\beta_1$ & 0.76 & 0.05\\
    $\beta_2$ & 0.04 & 0.04 \\
    $\theta$ & 0.57 & 0.19\\
    $\omega$ & -0.02 & 0.009\\ 
    \hline
\end{tabular}~
\end{table}

We fit the interaction model given in Equation \ref{e:influence-temporal-mod}, and fitted model parameter estimates are shown in Table \ref{tab:social.inf}. The estimate for $\omega$ is statistically significant, which shows that self-esteem has a moderating effect on social influence. The higher the self-esteem value, the smaller the impact friends' STEM interest has on one's own STEM interest at a later time.

This example illustrates how to accommodate a social influence moderator, and there are likely many other research contexts that would benefit from a moderation model. For example, social influence of risk taking behavior is likely moderated by the strength of relationships with family members; social influence of teacher beliefs is likely moderated by teachers' self-efficacy; and social influence on seeking mental health services may be moderated by gender and gender norms.

\subsection{Modeling Social Influence In Psychology}
We have given two toy examples of social influence among peers: how social influence impacts STEM interest and how self-esteem moderates the impact of social influence on STEM interest, but how can social influence models be used in other areas of Psychology? Existing literature investigating the impact of social influence often aggregate similarities among individuals rather than directly modeling social connections. We posit that collecting network data when investigating social influence can facilitate better social influence research with more nuanced understanding of the dynamics between relationships and behaviors. 

A social influence model can be used in any number of contexts. It can be used to quantify peer effects on various outcomes of interest whether they be academic or behavioral, given various types of relationships such as friendship, peer groups, co-studying, reading groups, etc. Social influence models that quantify influence and moderate influence with other variables can substantially contribute to the current landscape of social influence research.
Take risk-taking and defiant behavior as an example; social influence models not only allow us to understand to what extent risk-taking is influenced by peers, but also to identify the protective factors such as parent-child relationship health, attachment styles or discipline styles that keep a child resistant to social influence. This is not to claim that this research does not exist, but using social influence models to quantify influence or moderating influence effects is not common, and more prevalent use of social influence models may push current research into new directions. 

Social influence models can also be applied to information or disease networks, and these networks might be among individuals or various components of the body. Being able to address research questions - such as to what extent are social media followers, health care providers, journalists, etc. impacting individual beliefs and behaviors - would help researchers better understand information diffusion and design interventions. Current social network theory suggests that important or high status individuals are most influential in information diffusion, but many questions are left unanswered such as the role of individual variability in status and influence, which can help us better address critical societal problems as such the spread of misinformation.

Other disciplines may benefit from modeling social influence as well. For example, 
influence models can help us understand how consensus is reached among a group of individuals in terms of social cognition. These models are also useful in understanding the effectiveness of teams, moral reasoning, social inclusion or exclusion, as well as how teachers and other employees change their behaviors and beliefs. For example, \cite{zagenczyk2015social} found an association between workplace perceptions among individuals who shared advice-seeking ties but not friendship ties. They used a different method to estimate that association, but they could have used a social influence modeling approach. 

Finally, by including moderators and even mediators, researchers can answer a number of research questions about the heterogeneity of social influence. Our example about STEM interest only examined moderating effects on students that are being influenced, but similar models can be specified to understand and estimate the variability in how influential certain students are in the network. Students with many ties are commonly assumed to be influential, but influence can also be modeled as a function of covariates to identify which types of students are most influential. Because of the dyadic nature of networks and network relationships, a myriad of associations among variables can be investigated.

\section{Network Interference} 
In the presence of social influence, individual attributes may change as a direct result from individuals interacting. In experimental interventions or observational studies aimed at estimating a treatment effect, having study participants  influencing each other's outcomes is problematic. This is because the presence of social influence in these contexts violates something called the stable unit treatment value assumption \citep{rubin2005causal}, one of several causal inference assumptions needed for obtaining unbiased estimates of treatment effects. This assumption requires individual outcomes be impacted by their assigned treatment, not by the assigned treatment of other individuals, and this assumption is violated when social influence or information diffusion exists. This phenomenon of individuals impacting each other's outcomes is called network interference. 

Consider the following study by \cite{killen2022testing} in which third, fourth, and fifth grade classrooms were randomly assigned to participate in an intervention aimed at reducing prejudice and bias. In each school and for each grade, at least 1 classroom  was assigned to the treatment condition and at least 1 classroom was assigned to control. While there are many reasons to randomize within each school, one being to control for school-level effects, within-school randomization increases the likelihood of students in the treatment condition interacting with students in the control condition. 

When students in the treatment and control conditions interact, ideas and information from the treatment condition can be shared with students in the control condition. This particular study collected data on playing with a diverse set of peers as well as opinions about peer exclusion. In many elementary schools, children are segregated by class, but during recess and after-school activities, children are interacting with those in other classes. It is possible that as students in the treatment condition were choosing to diversify their play group and were becoming less likely to exclude kids of different racial backgrounds, their friends and peers in the control condition were witnessing and being influenced by these behaviors. While the research team found significant treatment effects, it is possible that in the absence of network interference, the intervention effects would have been even larger. 


Therefore, we believe that network interference should be considered by researchers. Although there is a robust literature about network interference (e.g. see \cite{an2018causal, forastiere2020identification, eckles2016design}), there are relatively few sources on how to account for network interference in the social sciences. Thus for the time being, our recommendation is to qualitatively consider the impacts of potential network interference when designing studies.


\section{Co-evolution}

\subsection{Selection and Influence are Confounded}
Thus far, we have discussed social selection and social influence and their respective models as separate entities yet, selection and influence are difficult and sometimes impossible to disentangle. In fact, with cross-sectional data, they are confounded and cannot be separated \citep{shalizi2011homophily}. Even with longitudinal data, selection and influence cannot always be separated and precisely estimated. 

In terms of causal processes, selection can be considered the cause for a network tie. We think of social selection as the reason two individuals share the relationship of interest. Homophily, the fact that individuals who are similar are likely to have or form ties with each other, is a common selection cause. On the other hand, we can think of influence as the causal process of changing one's outcomes due to their network ties. Then why the confusion? Individuals do not form ties and change their behavior in discrete time steps; it's a continuous process. We use smoking behaviors and friendship, defined as spending free time together in-person as an example. Are people friends because they are both smokers or non-smokers? It's possible. People may meet each other and interact because of this shared behavior. On the other hand, if two people are friends and one of them starts smoking, their non-smoking friend might be influenced to start smoking; alternatively, when a smoker and a non-smoker become friends, the non-smoker may help the smoker to quit smoking. 

Therefore, if we only collect data at a single time point, we have information about smoking and friendship, but we do not know when people started smoking or when people started being friends. And yet both models may find significant effects of smoking on friendship and of friendship on smoking! In the same vein that we tell our students that correlation is not causation, significant estimates in social selection and social influence models do not guarantee selection or influence. And in the same vein that causal inference researchers rely on strong assumptions to make strong claims, so too do network scientists when concluding social selection or influence. 

In cross-sectional data, we tend to focus only on those attributes that do not change when estimating selection. For example, we can examine social selection among friendship networks based on racial homophily since racial identity largely does not change during the course of friendship. Attributes that do not or are unlikely to change during the course of a study include traits such as income, neighborhood, religion, job position, and medical history.  Assuming social influence with cross-sectional data is more difficult because the data are unable to give much support to the assumption that the outcome has changed from some earlier time point and that the network was the reason for this change.

One way to tease out social influence is to collect longitudinal data, collecting data on smoking at time 1 and then again at time 2. Using the network that existed at time 2 or between time 1 and 2, one can more easily assume social influence because of the temporal nature of causal processes. If the outcome variable has changed, social influence is now possible. This is not to say that social selection has been eliminated altogether; changes in smoking unrelated to network ties could have occurred and social selection created changes in the network.  Still, most selection studies will attempt to choose network ties that tend to be stable as opposed to those that would form during the course of the study. As an example, suppose we have a friendship network for 5th graders. We collect data on recess games played in the beginning of 5th grade and recess games played at the end of 5th grade and measure the network for friendships during 5th grade. By modeling the change in games, we are in essence controlling for any endogenous effects of games on network ties. Because friendship ties tend to be pretty stable, we may feel more comfortable assuming change in recess games is based on social influence.

The above scenario for social influence is not perfect; by only having a single social network, we are assuming that the network is stable and that ties due to social selection didn't form during the course of the study. One solution to fully estimate social influence in the presence of social selection is to have both longitudinal data for the network and for the outcome of interest and model selection and influence simultaneously. 

\subsection{Co-evolution Models}

Let us begin with an example of peer drinking behavior among university students by \cite{Reifman2006}. In this study, the authors used a longitudinal model and found that students with similar drinking behaviors were more likely to be friends at subsequent time points, and students were likely to change their drinking behavior to mirror that of their friends.  While longitudinal models are common for estimating simultaneous effects, it may not be ideal for network data due to the independence assumption that is violated by network ties. 

\begin{figure} [hbt!]
    \centering
    \includegraphics[width=.95\textwidth]{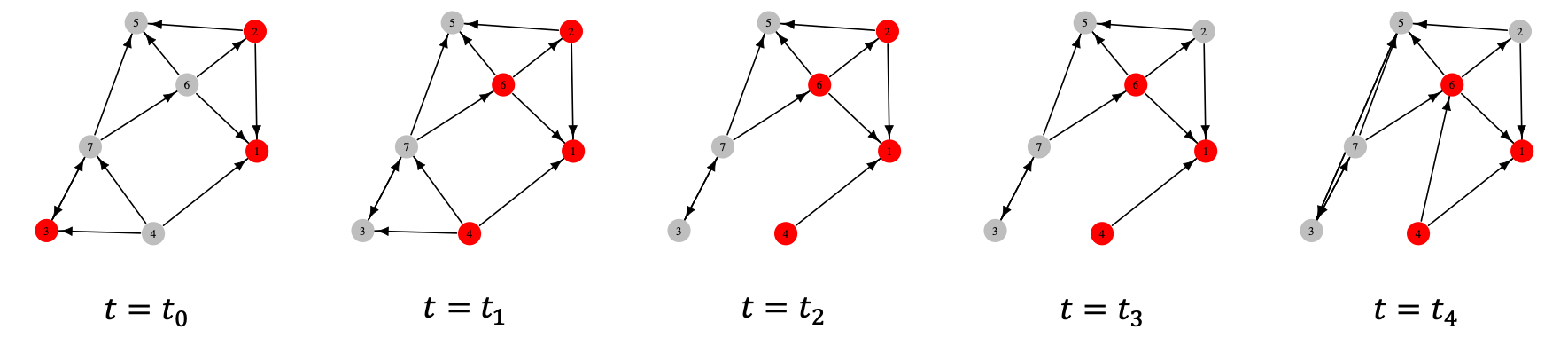}
    \caption{A simple example of the coevolution of network and node attributes over time shows how network ties impact a binary nodal attribute shown as gray or red; in turn, nodes that are a similar color are more likely to form future ties and nodes of different colors are more likely to dissolve ties over time.}
    \label{fig:coevolution}
\end{figure}

An appropriate social network model that models the co-evolution of selection and influence and accounts for network tie dependence in the selection model is called a co-evolution model \citep{Siena}. Figure \ref{fig:coevolution} shows a toy example of co-evolution with a network and a binary node attribute across four time points. At time $t=t_0$, we have a network of seven nodes whose binary attribute is given as red or gray, e.g., an indicator of whether students are binge-drinking on a weekly basis. 

Between time $t= t_0$ and time $t=t_1$, we see that influence has occurred and that node 3 has changed from red to gray and nodes 4 and 6 from gray to red. These attribute changes then result in changes to the network at time $t=t_2$, where the ties from node 4 to node 3 and node 4 to node 7 dissolve. Influence occurs again between $t_2$ and $t_3$ so that node 2 has changed from red to gray. Then the current state of node attributes result in the formation of ties from node 3 to node 5 and from node 4 to node 6 at time $t=t_4$. 

By observing temporal changes in both the network and node or dyad covariates, we are theoretically able to disentangle selection and influence. While several coevolution models have been proposed \citep{HeHoff2019, Gu2017, NIPS-Farajtabar}, the most popular method extends stochastic actor oriented models \citep{snijders1996stochastic} and is implemented in a software package called simulation investigation for empirical network analysis \citep[SIENA,][]{Siena}. 

In a stochastic actor oriented model, each node changes their out-going ties over time.  Nodes can dissolve existing ties they have with other nodes, they can initiate ties to nodes with whom they are not connected, or do nothing. The model assumes that every action happens separately as part of a continuous time process. Alternatively, a nodal attribute, such as smoking status or self-esteem, can change as a result of social influence. Social influence for all nodes is also happening such that attributes for a given node are changing as part of another continuous time process. The data can be considered snapshots of the state of the network and nodal attributes as discrete points in time. How these changes happen are part of the model estimation/simulation method; using the observed data, micro-simulations are carried out to optimize the parameters in the model (see \cite{snijders1996stochastic} for details). The model is fit based on which parameter values are most likely to result in the observed data, given the process specified by the model. 

Although SIENA models are complicated, its creators and users have been actively disseminating how to use these models, particularly in the psychological sciences. SIENA collaborators have produced an R package, online training modules, workshops, and even youtube videos. As a result, co-evolution models may be one of the most popular social network methods found in psychology research. In fact, SIENA models have been used to model the co-evolution of student friendship and academic achievement \citep{laninga2019role}; friendship and aggressive behaviors \citep{Zhang-SIENAEx}; friendship and social inclusion \citep{garrote2023friendship}; online peer relationships and ratings of stem-abilities and stem activities \citep{hopp2020supporting}; and
friendship networks and workplace bullying behavior ratings \citep{pauksztat2020targets}. See a systematic review of using SIENA to model peer influence and adolescence substance in \cite{henneberger2021peer}. 

It should be noted that SIENA software is not applicable to all network studies. SIENA requires longitudinal data of both the nodal attributes and the networks, minimal turnover in the network, and limited missing data. The challenge is therefore to conduct a study in which the individuals in the network do not change over time and the amount of missingness is trivial. Although some research suggests that SIENA handles missing data better than other methods \citep{ZandbergHuisman2019}, not all network studies are appropriate for SIENA.

\section{Ego Network Analysis}
Thus far, we have focused on methods for sociometric network data, which are network ties that have been collected among a group of individuals. For example, if we ask all students in a school who they like to play with, we can construct a sociometric network. Collecting sociometric network data is not always feasible, nor do all network research questions require a sociometric approach. An alternative is called ego network analysis. 

\begin{figure}[hbt!]
    \centering
    \includegraphics[width=.75\textwidth]{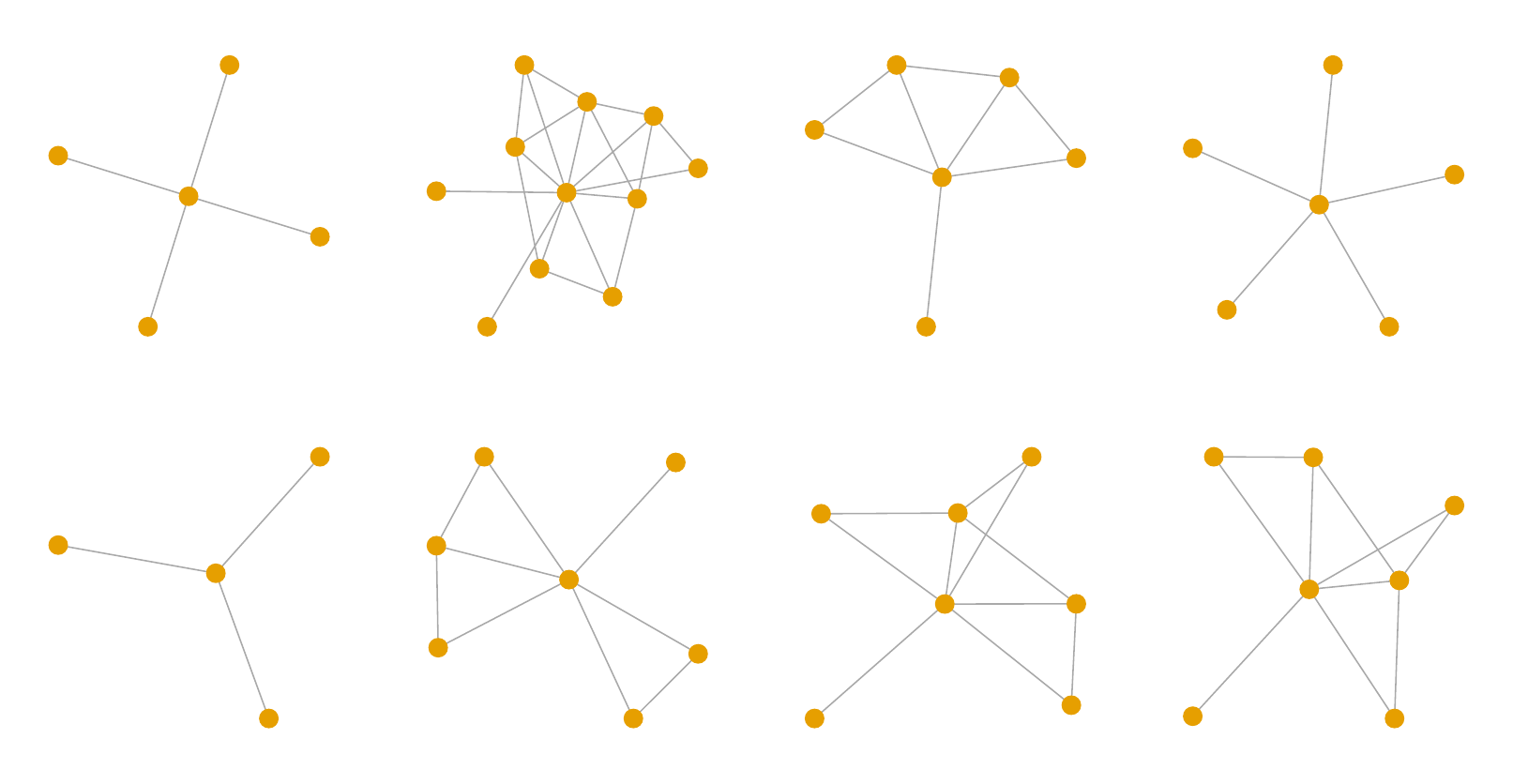}
    \caption{A example of an ego network dataset with 8 individual ego networks; each ego is connected to their alters and the alter connections are also included.}
    \label{f:egonetwork}
\end{figure}

An ego-centric or ego network is a network that starts with a single node, termed the ego. The network is constructed around the ego - for example, the ego lists their friends. The individuals nominated are called alters.  Many ego networks end here, but we generally consider an ego network to include the network ties among the alters. Ego networks also differ from sociometric networks in that network data from multiple egos constitute a single dataset. Figure \ref{f:egonetwork} gives an example of an ego network with 8 egos. Each ego is situated in their own network with their own alters, and the connections among alters are included. 

Research questions regarding ego networks often focus on the differences among the networks. Are egos with a certain attribute more or less likely to have specific network structures? Are egos with certain network characteristics more likely to have certain outcomes? Consider the following example from \cite{grier2016african}, who compared ego networks of Black students who participated in a specific affiliation group at a predominantly white university versus ego networks of Black students who did not participate in the affiliation group. They found that students in the group had higher numbers of alters and that higher proportions of their alters were same-race, same-gender, and same-university compared to non-participants. 

Furthermore, standard statistical methods can be applied to conduct ego network analysis. For example, \cite{grier2016african} used Kruskal–Wallis tests, commonly described as a non-parametric version of a t-test. Similarly, summaries of ego networks can be included in regression models as predictors or outcomes. Whereas sociometric network data are interdependent, ego networks of separate individuals can be assumed to be independent. There are some contexts where it would be more difficult to assume independence among ego networks, such as when an ego would be another ego's alter. In this case, the lack of independence in the statistical model is considered be a limitation; to our knowledge, it is not clear if specific methodology for inter-connected ego networks is needed.

\section{Network Psychometrics}
Network psychometrics consists of methods for the graphical or network representation of psychological constructs \citep{marsman2018introduction,van2015association,mcnally2015mental,fried2015loss,isvoranu2016network,kossakowski2016application,dalege2016toward,cramer2012dimensions}. Researchers working on network psychometrics argue that a mental disorder such as depression is better represented as a cluster of symptoms that directly influence one another rather than as a latent variable that underlies a constellation of symptoms \citep{marsman2018introduction}.  In this section, we briefly discuss network psychometrics with a focus on educating readers on what network psychometrics does and does not do.

Consider a survey of questionnaire items/variables regarding peer social exclusion. There might be 5 variables: moral reasoning, self-esteem, personal prejudice, intraversion, and school belonging. There might be 5-10 items purported to measure each construct.  

A mathematical representation of a network psychometric model can be written as
\begin{align}
    \text{Cor} (Y_j, Y_k| \boldsymbol{Y}^{-(j,k)}) = w_{jk}=w_{kj}, \label{networkp}
\end{align}where $Y_j$, $j=1,2,...,P$ represents  data on the $j$th behavior. 

The goal of this model is to identify a $P$ by $P$ relational structure among the $P$ items. Each item is conceptualized as a node in the network, and the presence of conditional dependence between variables is seen as an edge between two nodes. In Equation \ref{networkp}, $w_{jk}$ represents the weight of the edge between variables/nodes $j$ and $k$.

For our example, the goal of a network psychometric model would be to infer the connections among the 5 constructs, among the items within each construct, or both. We might find that there is evidence of a link/conditional dependence between self-esteem and moral reasoning and personal prejudice, but that intraversion and school belonging are not linked to the others. 

A key difference between a network psychometric model and a social network model is the goal of the analysis. In network psychometrics, we aim to construct the network of items and constructs, using responses by people. We assume that there is an underlying true dependence structure among these items. 
Network psychometric models are therefore measurement models aimed at mapping responses to a structural model, which is a network. It is therefore not surprising that the properties of a network psychometric model are often compared with those of a factor analytic model, a structural equation model, or an item response model.

These methods are in stark contrast to the methods introduced earlier. We assume that the network among a group of individuals is known and that the inference focuses on how or why certain individuals have ties (in social selection models) or  how the interactions from their network influence their own attributes (social influence). 

If we consider network psychometric models as measurement models, this brings up a separate issue, one that is a bit outside the scope of this manuscript and that is whether network psychometric models or standard measurement models are closer to the truth. For example, in demonstrating the statistical equivalence of an ising model and an item response model \citep{marsman2018introduction}, network psychometrics researchers considered correlated variables as the true data generating process of psychological outcomes instead of the latent psychological constructs. As a result, other researchers have drawn concerns over the application of network psychometric models \citep{neal2022critiques}. Latent variables are not just used to represent latent constructs; they also have longstanding roots in statistics as a way to account for various dependence structures as random effects. Is covariance structure a better way to understand and model psychological survey data than latent variables? It remains to be seen. 



On the other hand, if we have a network, regardless of what the nodes represent, we must consider whether networks inferred from network psychometric models are actually networks. That is, would any of the standard descriptions or analyses currently used for social network analysis apply? We urge caution because a network psychometric model is a Gaussian graphical model (GGM) that expresses the conditional dependencies among the random variables as a graph. The constructed networks follow certain covariance structures, and the values of the networks are bounded following the positive definiteness of the covariance matrices. Constructed covariance matrices do not possess the same properties as real-world relational networks, and thus the summaries of covariance matrices via summary statistics should not be interpreted in the same way as summaries of real-world networks. 

Ideas presented in network psychometrics have gained widespread popularity, but are also heavily criticized \citep[e.g.,][]{bringmann2018don,fried2017moving,neal2022critiques}. Regardless of the debate, network psychometrics has certainly drawn attention to the potential contributions of network science in psychology.

\section{Discussion}
Many studies in psychology occur in the real world in which individuals interact with one another. These interactions are complex and worthy of notice; and incorporating network analysis into psychological research allows researchers to answer complex questions about individual behaviors, opinions, and beliefs. 

To that end, we have introduced common network analytical approaches, with a heavy focus on network methods most applicable for psychology. For example, social selection models allow researchers to quantify the extent to which node and dyad-level attributes are associated with specified relationships or interactions. These models can be used to determine why and how individuals interact. Social influence models allow researchers to quantify the amount of influence that is spread across individuals in a group. These models are particularly useful for studies of peer effects or understanding the extent to which a consensus is reached under peer influence. 

 In addition, we have demonstrated that social influence impacts all researchers who conduct research on groups of individuals that might interact. Individual outcomes may be impacted by those with whom they interact, which may affect treatment effect estimation. Given the confounding between social selection and influence, we summarized how co-evolution models with longitudinal data can be used to estimate social selection and social influence. We have also briefly introduced ego network analysis and network psychometrics, both of which incorporate networks in very different ways than the methods discussed earlier. Ego network analysis focuses on a collection of egos whose networks are compared. Network psychometrics aims to infer a network of items or constructs based on response patterns. 

Despite the plethora of network methods available to researchers, we acknowledge that employing these methods in psychology research is not necessarily easy or simple. Collecting social network data among children in public schools can be challenging, so the alternative is to ask teachers to construct the networks among children in their classrooms. Even if data collection is allowed, data cleaning and data analysis likely present an additional hurdle. We encourage researchers to both seek training opportunities for network analysis and collaborations with network analysts. 
As network analytical methods become more widely used in psychology, network methods will naturally improve and in turn broaden in their applicability and coverage. 
\newpage
 \bibliography{main.bib}

\begin{thebibliography}{96}
\newcommand{\enquote}[1]{``#1''}
\expandafter\ifx\csname natexlab\endcsname\relax\def\natexlab#1{#1}\fi

\bibitem[{Airoldi et~al.(2008)Airoldi, Blei, Fienberg, and
  Xing}]{airoldi2008mixed}
Airoldi, E.~M., Blei, D.~M., Fienberg, S.~E., and Xing, E.~P. (2008),
  \enquote{Mixed membership stochastic blockmodels,} \textit{Journal of Machine
  Learning Research}, 9, 1981--2014.

\bibitem[{An(2018)}]{an2018causal}
An, W. (2018), \enquote{Causal inference with networked treatment diffusion,}
  \textit{Sociological Methodology}, 48, 152--181.

\bibitem[{Antonoplis and John(2022)}]{antonoplis2022has}
Antonoplis, S. and John, O.~P. (2022), \enquote{Who has different-race friends,
  and does it depend on context? Openness (to other), but not agreeableness,
  predicts lower racial homophily in friendship networks.} \textit{Journal of
  Personality and Social Psychology}, 122, 894.

\bibitem[{Arroyo et~al.(2019)Arroyo, Athreya, Cape, Chen, Priebe, and
  Vogelstein}]{arroyo2019inference}
Arroyo, J., Athreya, A., Cape, J., Chen, G., Priebe, C.~E., and Vogelstein,
  J.~T. (2019), \enquote{Inference for multiple heterogeneous networks with a
  common invariant subspace,} \textit{arXiv preprint arXiv:1906.10026}.

\bibitem[{Bringmann and Eronen(2018)}]{bringmann2018don}
Bringmann, L.~F. and Eronen, M.~I. (2018), \enquote{Don’t blame the model:
  Reconsidering the network approach to psychopathology.} \textit{Psychological
  Review}, 125, 606.

\bibitem[{Carnevale et~al.(2022)Carnevale, Huang, Yam, and
  Wang}]{carnevale2022laughing}
Carnevale, J.~B., Huang, L., Yam, K.~C., and Wang, L. (2022), \enquote{Laughing
  with me or laughing at me? The differential effects of leader humor
  expressions on follower status and influence at work,} \textit{Journal of
  Organizational Behavior}, 43, 1153--1171.

\bibitem[{Clauset et~al.(2004)Clauset, Newman, and Moore}]{clauset2004finding}
Clauset, A., Newman, M.~E., and Moore, C. (2004), \enquote{Finding community
  structure in very large networks,} \textit{Physical review E}, 70, 066111.

\bibitem[{Cooc and Kim(2017)}]{cooc2017peer}
Cooc, N. and Kim, J.~S. (2017), \enquote{Peer influence on children's reading
  skills: A social network analysis of elementary school classrooms.}
  \textit{Journal of Educational Psychology}, 109, 727.

\bibitem[{Cramer et~al.(2012)Cramer, Van~der Sluis, Noordhof, Wichers,
  Geschwind, Aggen, Kendler, and Borsboom}]{cramer2012dimensions}
Cramer, A.~O., Van~der Sluis, S., Noordhof, A., Wichers, M., Geschwind, N.,
  Aggen, S.~H., Kendler, K.~S., and Borsboom, D. (2012), \enquote{Dimensions of
  normal personality as networks in search of equilibrium: You can't like
  parties if you don't like people,} \textit{European Journal of Personality},
  26, 414--431.

\bibitem[{Dabbs et~al.(2020)Dabbs, Adhikari, and Sweet}]{CIDpaper}
Dabbs, B., Adhikari, S., and Sweet, T. (2020), \enquote{Conditionally
  Independent Dyads (CID) network models: A latent variable approach to
  statistical social network analysis,} \textit{Social Networks}, 63, 122--133.

\bibitem[{Dalege et~al.(2016)Dalege, Borsboom, van Harreveld, van~den Berg,
  Conner, and van~der Maas}]{dalege2016toward}
Dalege, J., Borsboom, D., van Harreveld, F., van~den Berg, H., Conner, M., and
  van~der Maas, H.~L. (2016), \enquote{Toward a formalized account of
  attitudes: The Causal Attitude Network (CAN) model.} \textit{Psychological
  review}, 123, 2.

\bibitem[{Doreian(1989)}]{doreian1989network}
Doreian, P. (1989), \enquote{Network autocorrelation models: Problems and
  prospects,} \textit{Spatial statistics: Past, present, future}, 369--89.

\bibitem[{Dorff et~al.(2023)Dorff, Gallop, and Minhas}]{dorff2023network}
Dorff, C., Gallop, M., and Minhas, S. (2023), \enquote{Network competition and
  civilian targeting during civil conflict,} \textit{British Journal of
  Political Science}, 53, 441--459.

\bibitem[{Eckles et~al.(2016)Eckles, Karrer, and Ugander}]{eckles2016design}
Eckles, D., Karrer, B., and Ugander, J. (2016), \enquote{Design and analysis of
  experiments in networks: Reducing bias from interference,} \textit{Journal of
  Causal Inference}, 5.

\bibitem[{Farajtabar et~al.(2015)Farajtabar, Wang, Gomez~Rodriguez, Li, Zha,
  and Song}]{NIPS-Farajtabar}
Farajtabar, M., Wang, Y., Gomez~Rodriguez, M., Li, S., Zha, H., and Song, L.
  (2015), \enquote{COEVOLVE: A Joint Point Process Model for Information
  Diffusion and Network Co-evolution,} in \textit{Advances in Neural
  Information Processing Systems}, eds. Cortes, C., Lawrence, N., Lee, D.,
  Sugiyama, M., and Garnett, R., Curran Associates, Inc., vol.~28.

\bibitem[{Fleming and Alexander(2001)}]{fleming2001benefits}
Fleming, V.~M. and Alexander, J.~M. (2001), \enquote{The benefits of peer
  collaboration: A replication with a delayed posttest,} \textit{Contemporary
  Educational Psychology}, 26, 588--601.

\bibitem[{Forastiere et~al.(2020)Forastiere, Airoldi, and
  Mealli}]{forastiere2020identification}
Forastiere, L., Airoldi, E.~M., and Mealli, F. (2020), \enquote{Identification
  and estimation of treatment and interference effects in observational studies
  on networks,} \textit{Journal of the American Statistical Association},
  1--18.

\bibitem[{Frank et~al.(2014)Frank, Lo, and Sun}]{frank2014social}
Frank, K.~A., Lo, Y.-J., and Sun, M. (2014), \enquote{Social network analysis
  of the influences of educational reforms on teachers practices and
  interactions,} \textit{Zeitschrift f{\"u}r Erziehungswissenschaft}, 17,
  117--134.

\bibitem[{Frank and Strauss(1986)}]{FrankStrauss1986}
Frank, O. and Strauss, D. (1986), \enquote{Markov Graphs,} \textit{Journal of
  the American Statistical Association}, 81, 832--842.

\bibitem[{Fried et~al.(2015)Fried, Bockting, Arjadi, Borsboom, Amshoff, Cramer,
  Epskamp, Tuerlinckx, Carr, and Stroebe}]{fried2015loss}
Fried, E.~I., Bockting, C., Arjadi, R., Borsboom, D., Amshoff, M., Cramer,
  A.~O., Epskamp, S., Tuerlinckx, F., Carr, D., and Stroebe, M. (2015),
  \enquote{From loss to loneliness: The relationship between bereavement and
  depressive symptoms.} \textit{Journal of abnormal psychology}, 124, 256.

\bibitem[{Fried and Cramer(2017)}]{fried2017moving}
Fried, E.~I. and Cramer, A.~O. (2017), \enquote{Moving forward: challenges and
  directions for psychopathological network theory and methodology,}
  \textit{Perspectives on Psychological Science}, 12, 999--1020.

\bibitem[{Friel et~al.(2016)Friel, Rastelli, Wyse, and
  Raftery}]{friel2016interlocking}
Friel, N., Rastelli, R., Wyse, J., and Raftery, A.~E. (2016),
  \enquote{Interlocking directorates in Irish companies using a latent space
  model for bipartite networks,} \textit{Proceedings of the National Academy of
  Sciences}, 113, 6629--6634.

\bibitem[{Garrote et~al.(2023)Garrote, Zurbriggen, and
  Schwab}]{garrote2023friendship}
Garrote, A., Zurbriggen, C.~L., and Schwab, S. (2023), \enquote{Friendship
  networks in inclusive elementary classrooms: Changes and stability related to
  students’ gender and self-perceived social inclusion,} \textit{Social
  Psychology of Education}, 1--19.

\bibitem[{Gilbert(1959)}]{gilbert1959random}
Gilbert, E.~N. (1959), \enquote{Random graphs,} \textit{The Annals of
  Mathematical Statistics}, 30, 1141--1144.

\bibitem[{Girvan and Newman(2002)}]{girvan2002community}
Girvan, M. and Newman, M.~E. (2002), \enquote{Community structure in social and
  biological networks,} \textit{Proceedings of the national academy of
  sciences}, 99, 7821--7826.

\bibitem[{Gollini and Murphy(2016)}]{gollini2016joint}
Gollini, I. and Murphy, T.~B. (2016), \enquote{Joint modeling of multiple
  network views,} \textit{Journal of Computational and Graphical Statistics},
  25, 246--265.

\bibitem[{Grier-Reed and Wilson(2016)}]{grier2016african}
Grier-Reed, T. and Wilson, R.~J. (2016), \enquote{The African American Student
  Network: An exploration of Black students’ ego networks at a predominantly
  White institution,} \textit{Journal of Black Psychology}, 42, 374--386.

\bibitem[{Gu and Yu(2022)}]{gu2022joint}
Gu, J. and Yu, P.~L. (2022), \enquote{Joint latent space models for ranking
  data and social network,} \textit{Statistics and Computing}, 32, 51.

\bibitem[{Gu et~al.(2017)Gu, Sun, and Gao}]{Gu2017}
Gu, Y., Sun, Y., and Gao, J. (2017), \enquote{The Co-Evolution Model for Social
  Network Evolving and Opinion Migration,} in \textit{Proceedings of the 23rd
  ACM SIGKDD International Conference on Knowledge Discovery and Data Mining},
  New York, NY, USA: Association for Computing Machinery, KDD '17, p.
  175–184.

\bibitem[{Hamamoto et~al.(2021)Hamamoto, Mizobata, Ishikawa, and
  Itakura}]{hamamoto2021examining}
Hamamoto, H., Mizobata, R., Ishikawa, M., and Itakura, S. (2021),
  \enquote{Examining the social influence of reputation for partner
  productivity level on the collaborative task performance of young children,}
  \textit{Infant and Child Development}, 30, e2213.

\bibitem[{Handcock(2003)}]{handcock2003assessing}
Handcock, M. (2003), \enquote{Assessing degeneracy in statistical models of
  social networks,} Tech. rep., Working Paper No. 39, Center for Statistics and
  the Social Sciences, University of Washington, Seattle, WA.

\bibitem[{Hanneke et~al.(2010)Hanneke, Fu, and Xing}]{HannekeFu2010}
Hanneke, S., Fu, W., and Xing, E.~P. (2010), \enquote{Discrete temporal models
  of social networks,} \textit{Electronic Journal of Statistics}, 4, 585--605.

\bibitem[{Harris et~al.(2019)Harris, Halpern, Whitsel, Hussey, Killeya-Jones,
  Tabor, and Dean}]{harris2019cohort}
Harris, K.~M., Halpern, C.~T., Whitsel, E.~A., Hussey, J.~M., Killeya-Jones,
  L.~A., Tabor, J., and Dean, S.~C. (2019), \enquote{Cohort profile: The
  national longitudinal study of adolescent to adult health (add health),}
  \textit{International journal of epidemiology}, 48, 1415--1415k.

\bibitem[{He and Hoff(2019)}]{HeHoff2019}
He, Y. and Hoff, P.~D. (2019), \enquote{Multiplicative coevolution regression
  models for longitudinal networks and nodal attributes,} \textit{Social
  Networks}, 57, 54--62.

\bibitem[{Henneberger et~al.(2021)Henneberger, Mushonga, and
  Preston}]{henneberger2021peer}
Henneberger, A.~K., Mushonga, D.~R., and Preston, A.~M. (2021), \enquote{Peer
  influence and adolescent substance use: A systematic review of dynamic social
  network research,} \textit{Adolescent Research Review}, 6, 57--73.

\bibitem[{Hoff(2021)}]{hoff2021additive}
Hoff, P. (2021), \enquote{Additive and multiplicative effects network models,}
  \textit{Statistical Science}, 36, 34--50.

\bibitem[{Hoff(2005)}]{hoff2005bilinear}
Hoff, P.~D. (2005), \enquote{Bilinear mixed-effects models for dyadic data,}
  \textit{Journal of the american Statistical association}, 100, 286--295.

\bibitem[{Hoff et~al.(2002)Hoff, Raftery, and Handcock}]{hoff2002latent}
Hoff, P.~D., Raftery, A.~E., and Handcock, M.~S. (2002), \enquote{Latent space
  approaches to social network analysis,} \textit{Journal of the american
  Statistical association}, 97, 1090--1098.

\bibitem[{Holland et~al.(1983)Holland, Laskey, and
  Leinhardt}]{holland1983stochastic}
Holland, P.~W., Laskey, K.~B., and Leinhardt, S. (1983), \enquote{Stochastic
  blockmodels: First steps,} \textit{Social networks}, 5, 109--137.

\bibitem[{Hopp et~al.(2020)Hopp, Stoeger, and Ziegler}]{hopp2020supporting}
Hopp, M.~D., Stoeger, H., and Ziegler, A. (2020), \enquote{The supporting role
  of mentees’ peers in online mentoring: a longitudinal social network
  analysis of peer influence,} \textit{Frontiers in Psychology}, 11, 1929.

\bibitem[{Isvoranu et~al.(2016)Isvoranu, Borsboom, van Os, and
  Guloksuz}]{isvoranu2016network}
Isvoranu, A.-M., Borsboom, D., van Os, J., and Guloksuz, S. (2016), \enquote{A
  network approach to environmental impact in psychotic disorder: brief
  theoretical framework,} \textit{Schizophrenia Bulletin}, 42, 870--873.

\bibitem[{Killen et~al.(2022)Killen, Burkholder, D'Esterre, Sims, Glidden, Yee,
  Luken~Raz, Elenbaas, Rizzo, Woodward, et~al.}]{killen2022testing}
Killen, M., Burkholder, A.~R., D'Esterre, A.~P., Sims, R.~N., Glidden, J., Yee,
  K.~M., Luken~Raz, K.~V., Elenbaas, L., Rizzo, M.~T., Woodward, B., et~al.
  (2022), \enquote{Testing the effectiveness of the Developing Inclusive Youth
  program: A multisite randomized control trial,} \textit{Child development},
  93, 732--750.

\bibitem[{Kim and Schallert(2014)}]{KIM2014134}
Kim, T. and Schallert, D.~L. (2014), \enquote{Mediating effects of teacher
  enthusiasm and peer enthusiasm on students' interest in the college
  classroom,} \textit{Contemporary Educational Psychology}, 39, 134--144.

\bibitem[{Kossakowski et~al.(2016)Kossakowski, Epskamp, Kieffer, van Borkulo,
  Rhemtulla, and Borsboom}]{kossakowski2016application}
Kossakowski, J.~J., Epskamp, S., Kieffer, J.~M., van Borkulo, C.~D., Rhemtulla,
  M., and Borsboom, D. (2016), \enquote{The application of a network approach
  to Health-Related Quality of Life (HRQoL): introducing a new method for
  assessing HRQoL in healthy adults and cancer patients,} \textit{Quality of
  Life Research}, 25, 781--792.

\bibitem[{Laninga-Wijnen et~al.(2019)Laninga-Wijnen, Gremmen, Dijkstra,
  Veenstra, Vollebergh, and Harakeh}]{laninga2019role}
Laninga-Wijnen, L., Gremmen, M.~C., Dijkstra, J.~K., Veenstra, R., Vollebergh,
  W.~A., and Harakeh, Z. (2019), \enquote{The role of academic status norms in
  friendship selection and influence processes related to academic
  achievement.} \textit{Developmental Psychology}, 55, 337.

\bibitem[{Lazega and Snijders(2015)}]{lazega2015multilevel}
Lazega, E. and Snijders, T.~A. (2015), \textit{Multilevel network analysis for
  the social sciences: Theory, methods and applications}, vol.~12, Springer.

\bibitem[{Leblanc et~al.(2022)Leblanc, Rousseau, and
  Harvey}]{leblanc2022leader}
Leblanc, P.-M., Rousseau, V., and Harvey, J.-F. (2022), \enquote{Leader
  humility and team innovation: The role of team reflexivity and team proactive
  personality,} \textit{Journal of Organizational Behavior}, 43, 1396--1409.

\bibitem[{Leenders(2002)}]{leenders2002modeling}
Leenders, R. T.~A. (2002), \enquote{Modeling social influence through network
  autocorrelation: constructing the weight matrix,} \textit{Social Networks},
  24, 21--47.

\bibitem[{Leung et~al.(2022)Leung, Piera Pi-Sunyer, Ahmed, Foulkes, Griffin,
  Sakhardande, Bennett, Dunning, Griffiths, Parker,
  et~al.}]{leung2022susceptibility}
Leung, J.~T., Piera Pi-Sunyer, B., Ahmed, S.~P., Foulkes, L., Griffin, C.,
  Sakhardande, A., Bennett, M., Dunning, D.~L., Griffiths, K., Parker, J.,
  et~al. (2022), \enquote{Susceptibility to prosocial and antisocial influence
  in adolescence following mindfulness training,} \textit{Infant and Child
  Development}, e2386.

\bibitem[{Liu et~al.(2021)Liu, Zhang, and Dunson}]{liu2021graph}
Liu, M., Zhang, Z., and Dunson, D.~B. (2021), \enquote{Graph auto-encoding
  brain networks with applications to analyzing large-scale brain imaging
  datasets,} \textit{Neuroimage}, 245, 118750.

\bibitem[{Marsman et~al.(2018)Marsman, Borsboom, Kruis, Epskamp, van Bork,
  Waldorp, Maas, and Maris}]{marsman2018introduction}
Marsman, M., Borsboom, D., Kruis, J., Epskamp, S., van Bork, R., Waldorp, L.,
  Maas, H. v.~d., and Maris, G. (2018), \enquote{An introduction to network
  psychometrics: Relating ising network models to item response theory models,}
  \textit{Multivariate behavioral research}, 53, 15--35.

\bibitem[{McGill et~al.(2012)McGill, Way, and Hughes}]{mcgill2012intra}
McGill, R.~K., Way, N., and Hughes, D. (2012), \enquote{Intra-and interracial
  best friendships during middle school: Links to social and emotional
  well-being,} \textit{Journal of Research on Adolescence}, 22, 722--738.

\bibitem[{McNally et~al.(2015)McNally, Robinaugh, Wu, Wang, Deserno, and
  Borsboom}]{mcnally2015mental}
McNally, R.~J., Robinaugh, D.~J., Wu, G.~W., Wang, L., Deserno, M.~K., and
  Borsboom, D. (2015), \enquote{Mental disorders as causal systems: A network
  approach to posttraumatic stress disorder,} \textit{Clinical Psychological
  Science}, 3, 836--849.

\bibitem[{Minhas et~al.(2016)Minhas, Hoff, and Ward}]{minhas2016new}
Minhas, S., Hoff, P.~D., and Ward, M.~D. (2016), \enquote{A new approach to
  analyzing coevolving longitudinal networks in international relations,}
  \textit{Journal of Peace Research}, 53, 491--505.

\bibitem[{Minhas et~al.(2019)Minhas, Hoff, and Ward}]{minhas2019inferential}
--- (2019), \enquote{Inferential Approaches for Network Analysis: AMEN for
  Latent Factor Models,} \textit{Political Analysis}, 27, 208--222.

\bibitem[{Molloy et~al.(2011)Molloy, Gest, and Rulison}]{molloy2011peer}
Molloy, L.~E., Gest, S.~D., and Rulison, K.~L. (2011), \enquote{Peer influences
  on academic motivation: Exploring multiple methods of assessing youths’
  most “influential” peer relationships,} \textit{The Journal of Early
  Adolescence}, 31, 13--40.

\bibitem[{Neal et~al.(2022)Neal, Forbes, Neal, Brusco, Krueger, Markon,
  Steinley, Wasserman, and Wright}]{neal2022critiques}
Neal, Z.~P., Forbes, M.~K., Neal, J.~W., Brusco, M.~J., Krueger, R., Markon,
  K., Steinley, D., Wasserman, S., and Wright, A.~G. (2022), \enquote{Critiques
  of network analysis of multivariate data in psychological science,}
  \textit{Nature Reviews Methods Primers}, 2, 90.

\bibitem[{Newman and Girvan(2004)}]{newman2004finding}
Newman, M.~E. and Girvan, M. (2004), \enquote{Finding and evaluating community
  structure in networks,} \textit{Physical review E}, 69, 026113.

\bibitem[{Paluck et~al.(2016)Paluck, Shepherd, and Aronow}]{paluck2016changing}
Paluck, E.~L., Shepherd, H., and Aronow, P.~M. (2016), \enquote{Changing
  climates of conflict: A social network experiment in 56 schools,}
  \textit{Proceedings of the National Academy of Sciences}, 113, 566--571.

\bibitem[{Pauksztat and Salin(2020)}]{pauksztat2020targets}
Pauksztat, B. and Salin, D. (2020), \enquote{Targets' social relationships as
  antecedents and consequences of workplace bullying: a social network
  perspective,} \textit{Frontiers in psychology}, 10, 3077.

\bibitem[{Pons and Latapy(2006)}]{pons2006computing}
Pons, P. and Latapy, M. (2006), \enquote{Computing communities in large
  networks using random walks,} \textit{Journal of graph algorithms and
  applications}, 10, 191--218.

\bibitem[{Reifman et~al.(2006)Reifman, Watson, and McCourt}]{Reifman2006}
Reifman, A., Watson, W.~K., and McCourt, A. (2006), \enquote{Social Networks
  and College Drinking: Probing Processes of Social Influence and Selection,}
  \textit{Personality and Social Psychology Bulletin}, 32, 820--832, pMID:
  16648206.

\bibitem[{Ripley et~al.(2011)Ripley, Snijders, Boda, V{\"o}r{\"o}s, and
  Preciado}]{Siena}
Ripley, R.~M., Snijders, T.~A., Boda, Z., V{\"o}r{\"o}s, A., and Preciado, P.
  (2011), \enquote{Manual for SIENA version 4.0,} \textit{University of
  Oxford}.

\bibitem[{Risi et~al.(2003)Risi, Gerhardstein, and Kistner}]{risi2003children}
Risi, S., Gerhardstein, R., and Kistner, J. (2003), \enquote{Children's
  classroom peer relationships and subsequent educational outcomes,}
  \textit{Journal of Clinical Child and Adolescent Psychology}, 32, 351--361.

\bibitem[{Robins et~al.(2001)Robins, Pattison, and Elliott}]{robins2001network}
Robins, G., Pattison, P., and Elliott, P. (2001), \enquote{Network models for
  social influence processes,} \textit{Psychometrika}, 66, 161--189.

\bibitem[{Robins et~al.(2007)Robins, Pattison, Kalish, and
  Lusher}]{robins2007introduction}
Robins, G., Pattison, P., Kalish, Y., and Lusher, D. (2007), \enquote{An
  introduction to exponential random graph (p*) models for social networks,}
  \textit{Social networks}, 29, 173--191.

\bibitem[{Rubin(2005)}]{rubin2005causal}
Rubin, D.~B. (2005), \enquote{Causal inference using potential outcomes:
  Design, modeling, decisions,} \textit{Journal of the American Statistical
  Association}, 100, 322--331.

\bibitem[{Saleh et~al.(2007)Saleh, Lazonder, and de~Jong}]{saleh2007}
Saleh, M., Lazonder, A.~W., and de~Jong, T. (2007), \enquote{Structuring
  collaboration in mixed-ability groups to promote verbal interaction,
  learning, and motivation of average-ability students,} \textit{Contemporary
  Educational Psychology}, 32, 314--331.

\bibitem[{Salter-Townshend and McCormick(2017)}]{salter2017latent}
Salter-Townshend, M. and McCormick, T.~H. (2017), \enquote{Latent space models
  for multiview network data,} \textit{The annals of applied statistics}, 11,
  1217.

\bibitem[{Sarkar et~al.(2007)Sarkar, Siddiqi, and Gordon}]{sarkar2007latent}
Sarkar, P., Siddiqi, S.~M., and Gordon, G.~J. (2007), \enquote{A latent space
  approach to dynamic embedding of co-occurrence data,} in \textit{Artificial
  Intelligence and Statistics}, PMLR, pp. 420--427.

\bibitem[{Sewell and Chen(2015)}]{sewell2015latent}
Sewell, D.~K. and Chen, Y. (2015), \enquote{Latent space models for dynamic
  networks,} \textit{Journal of the american statistical association}, 110,
  1646--1657.

\bibitem[{Shalizi and Thomas(2011)}]{shalizi2011homophily}
Shalizi, C. and Thomas, A. (2011), \enquote{Homophily and contagion are
  generically confounded in observational social network studies,}
  \textit{Sociological Methods \& Research}, 40, 211--239.

\bibitem[{Siew(2013)}]{siew2013community}
Siew, C.~S. (2013), \enquote{Community structure in the phonological network,}
  \textit{Frontiers in psychology}, 4, 553.

\bibitem[{Skinner-Dorkenoo et~al.(2023)Skinner-Dorkenoo, George, Wages~III,
  S{\'a}nchez, and Perry}]{skinner2023systemic}
Skinner-Dorkenoo, A.~L., George, M., Wages~III, J.~E., S{\'a}nchez, S., and
  Perry, S.~P. (2023), \enquote{A systemic approach to the psychology of racial
  bias within individuals and society,} \textit{Nature Reviews Psychology}, 2,
  392--406.

\bibitem[{Slaughter and Koehly(2016)}]{slaughter2016multilevel}
Slaughter, A.~J. and Koehly, L.~M. (2016), \enquote{Multilevel models for
  social networks: Hierarchical Bayesian approaches to exponential random graph
  modeling,} \textit{Social Networks}, 44, 334--345.

\bibitem[{Snijders(1996)}]{snijders1996stochastic}
Snijders, T. (1996), \enquote{Stochastic actor-oriented models for network
  change,} \textit{The Journal of Mathematical Sociology}, 21, 149--172.

\bibitem[{Snijders et~al.(2006)Snijders, Pattison, Robins, and
  Handcock}]{Snijders2006ergms}
Snijders, T. A.~B., Pattison, P.~E., Robins, G.~L., and Handcock, M.~S. (2006),
  \enquote{New Specification For Exponential Random Graph Models,}
  \textit{Sociological Methodology}, 36, 99--153.

\bibitem[{Spillane et~al.(2018)Spillane, Hopkins, and
  Sweet}]{spillanehopkinssweetBELIEFS}
Spillane, J.~P., Hopkins, M., and Sweet, T. (2018), \enquote{School District
  Educational Infrastructure and Change at Scale: Teacher Peer Interactions and
  Their Beliefs About Mathematics Instruction,} \textit{American Educational
  Research Journal}, 55, 532--571, dOI: 10.3102/0002831217743928.

\bibitem[{Stevens et~al.(2012)Stevens, Tappon, Garg, and
  Fair}]{stevens2012functional}
Stevens, A.~A., Tappon, S.~C., Garg, A., and Fair, D.~A. (2012),
  \enquote{Functional brain network modularity captures inter-and
  intra-individual variation in working memory capacity,} \textit{PloS one}, 7,
  e30468.

\bibitem[{Stivala et~al.(2020)Stivala, Robins, and
  Lomi}]{stivala2020exponential}
Stivala, A., Robins, G., and Lomi, A. (2020), \enquote{Exponential random graph
  model parameter estimation for very large directed networks,} \textit{PloS
  one}, 15, e0227804.

\bibitem[{Sweet and Adhikari(2020)}]{Sweet2020Influence}
Sweet, T. and Adhikari, S. (2020), \enquote{A Latent Space Network Model for
  Social Influence,} \textit{Psychometrika}, 85, 251--274.

\bibitem[{Sweet et~al.(2013)Sweet, Thomas, and Junker}]{sweet2013jebs}
Sweet, T.~M., Thomas, A.~C., and Junker, B.~W. (2013), \enquote{Hierarchical
  Network Models for Education Research: Hierarchical Latent Space Models,}
  \textit{Journal of Educational and Behavioral Statistics}, 38, 295--318.

\bibitem[{Valente(2005)}]{valente2005network}
Valente, T.~W. (2005), \enquote{Network models and methods for studying the
  diffusion of innovations,} in \textit{Models and methods in social network
  analysis}, eds. Carrington, P., Scott, J., and Wasserman, S., New York, NY:
  Cambridge University Press, pp. 98--116.

\bibitem[{van Borkulo et~al.(2015)van Borkulo, Boschloo, Borsboom, Penninx,
  Waldorp, and Schoevers}]{van2015association}
van Borkulo, C., Boschloo, L., Borsboom, D., Penninx, B.~W., Waldorp, L.~J.,
  and Schoevers, R.~A. (2015), \enquote{Association of symptom network
  structure with the course of depression,} \textit{JAMA psychiatry}, 72,
  1219--1226.

\bibitem[{Verdecho et~al.(2011)Verdecho, Alfaro-Saiz, and
  Rodr{\'\i}guez-Rodr{\'\i}guez}]{verdecho2011review}
Verdecho, M.-J., Alfaro-Saiz, J.-J., and Rodr{\'\i}guez-Rodr{\'\i}guez, R.
  (2011), \enquote{A Review of factors influencing collaborative
  relationships,} in \textit{Adaptation and Value Creating Collaborative
  Networks: 12th IFIP WG 5.5 Working Conference on Virtual Enterprises, PRO-VE
  2011, S{\~a}o Paulo, Brazil, October 17-19, 2011. Proceedings 12}, Springer,
  pp. 535--542.

\bibitem[{Wang et~al.(2021)Wang, Lin, Cole, and Zhang}]{wang2021learning}
Wang, L., Lin, F.~V., Cole, M., and Zhang, Z. (2021), \enquote{Learning clique
  subgraphs in structural brain network classification with application to
  crystallized cognition,} \textit{NeuroImage}, 225, 117493.

\bibitem[{Wang et~al.(2019)Wang, Zhang, and Dunson}]{wang2019symmetric}
Wang, L., Zhang, Z., and Dunson, D. (2019), \enquote{Symmetric bilinear
  regression for signal subgraph estimation,} \textit{IEEE Transactions on
  Signal Processing}, 67, 1929--1940.

\bibitem[{Wang et~al.(2013)Wang, Robins, Pattison, and
  Lazega}]{wang2013exponential}
Wang, P., Robins, G., Pattison, P., and Lazega, E. (2013), \enquote{Exponential
  random graph models for multilevel networks,} \textit{Social networks}, 35,
  96--115.

\bibitem[{Wang and Edgerton(2022)}]{wang2022resilience}
Wang, S. and Edgerton, J. (2022), \enquote{Resilience to stress in bipartite
  networks: application to the Islamic State recruitment network,}
  \textit{Journal of Complex Networks}, 10, cnac017.

\bibitem[{Wang et~al.(2023)Wang, Paul, and De~Boeck}]{wang2023joint}
Wang, S., Paul, S., and De~Boeck, P. (2023), \enquote{Joint latent space model
  for social networks with multivariate attributes,} \textit{Psychometrika},
  88, 1197--1227.

\bibitem[{Wasserman et~al.(1994)Wasserman, Faust, et~al.}]{wasserman1994social}
Wasserman, S., Faust, K., et~al. (1994), \textit{Social network analysis:
  Methods and applications}, vol.~8, Cambridge university press.

\bibitem[{Wasserman and Pattison(1996)}]{wasserman1996logit}
Wasserman, S. and Pattison, P. (1996), \enquote{Logit models and logistic
  regressions for social networks: I. An introduction to Markov graphs andp,}
  \textit{Psychometrika}, 61, 401--425.

\bibitem[{White et~al.(2021)White, Russell, Qualter, Owens, and
  Psychogiou}]{white2021peer}
White, R., Russell, G., Qualter, P., Owens, M., and Psychogiou, L. (2021),
  \enquote{Do peer relationships mediate the association between children’s
  facial emotion recognition ability and their academic attainment? Findings
  from the ALSPAC study,} \textit{Contemporary Educational Psychology}, 64,
  101942.

\bibitem[{Zagenczyk et~al.(2015)Zagenczyk, Purvis, Shoss, Scott, and
  Cruz}]{zagenczyk2015social}
Zagenczyk, T.~J., Purvis, R.~L., Shoss, M.~K., Scott, K.~L., and Cruz, K.~S.
  (2015), \enquote{Social influence and leader perceptions: Multiplex social
  network ties and similarity in leader--member exchange,} \textit{Journal of
  Business and Psychology}, 30, 105--117.

\bibitem[{Zandberg and Huisman(2019)}]{ZandbergHuisman2019}
Zandberg, T. and Huisman, M. (2019), \enquote{Missing behavior data in
  longitudinal network studies: the impact of treatment methods on estimated
  effect parameters in stochastic actor oriented models,} \textit{Social
  Network Analysis and Mining}, 9, 8.

\bibitem[{Zhang et~al.(2020)Zhang, Liu, and Zhang}]{Zhang-SIENAEx}
Zhang, M., Liu, H., and Zhang, Y. (2020), \enquote{Adolescent Social Networks
  and Physical, Verbal, and Indirect Aggression in China: The Moderating Role
  of Gender,} \textit{Frontiers in Psychology}, 11.

\end{thebibliography}
 \bibliographystyle{asa2}
\end{document}